\documentclass[aoas,MSNbibl,nameyear,seceqn,dvips]{arximspdf}
\usepackage[mathscr]{eucal}
\usepackage{dcolumn}
\usepackage{graphicx}
\doi{10.1214/12-AOAS617} 
\volume{7}
\issue{2}
\pubyear{2013}
\firstpage{1010}
\lastpage{1039}

\makeatletter

\newcolumntype{d}[1]{D{.}{.}{#1}}

\newcommand{\ML}{\mathrm{ML}}
\newcommand{\LB}{\mathrm{LB}}

\newcommand{\mY}{\mathscr{Y}}
\newcommand{\bA}{\mathbf{A}}
\newcommand{\bx}{\mathbf{x}}
\newcommand{\bg}{\mathbf{g}}
\newcommand{\bt}{\mathbf{g}^\star}
\newcommand{\bY}{\mathbf{Y}}
\newcommand{\mZ}{\mathscr{Z}}
\newcommand{\mD}{\mathscr{D}}
\newcommand{\bZ}{\mathbf{Z}}

\newcommand{\bta}{\bolds{\eta}}
\newcommand{\bmu}{\bolds{\mu}}
\newcommand{\btheta}{\bolds{\theta}}
\newcommand{\bgamma}{\bolds{\gamma}}
\newcommand{\balpha}{\bolds{\alpha}}
\newcommand{\bpi}{\bolds{\pi}}

\newcommand{\msim}[1]{\stackrel{#1}{\sim}}
\newcommand{\given}{\mid}

\makeatother

\begin{document}
\begin{frontmatter}

\title{Model-based clustering of large networks\thanksref{T1}}
\runtitle{Model-based clustering of large networks}

\thankstext{T1}{Supported in part by the Office of Naval Research (ONR Grant
N00014-08-1-1015) and the National Institutes of Health (NIH Grant 1
R01 GM083603). Experiments in this work were also supported in part
through instrumentation funded by the National Science Foundation (NSF
Grant OCI-0821527).}

\begin{aug}
\author[A]{\fnms{Duy Q.} \snm{Vu}\ead[label=e1]{duy.vu@unimelb.edu.au}},
\author[B]{\fnms{David R.} \snm{Hunter}\corref{}\ead[label=e2]{dhunter@stat.psu.edu}}
\and
\author[C]{\fnms{Michael} \snm{Schweinberger}}
\runauthor{D. Q. Vu, D. R. Hunter and M. Schweinberger}
\affiliation{University of Melbourne, Pennsylvania State
University and Rice~University}
\address[A]{D. Q. Vu\\
Department of Mathematics\\
\quad and Statistics\\
University of Melbourne\\
Parkville, Victoria 3010\\
Australia\\
\printead{e1}}
\address[B]{D. R. Hunter\\
Department of Statistics\\
Pennsylvania State University\\
University Park, Pennsylvania 16802\\
USA\\
\printead{e2}} 
\address[C]{M. Schweinberger\\
Department of Statistics\\
Rice University\\
MS 138 P.O. Box 1892\\
Houston, Texas 77251-1892\\
USA}
\end{aug}

\received{\smonth{5} \syear{2012}}
\revised{\smonth{11} \syear{2012}}

%
\begin{abstract}
We describe a network clustering framework, based on finite mixture
models, that can be applied to discrete-valued networks with hundreds
of thousands of nodes and billions of edge variables. Relative to other
recent model-based clustering work for networks, we introduce a more
flexible modeling framework, improve the variational-approximation
estimation algorithm, discuss and implement standard error estimation
via a parametric bootstrap approach, and apply these methods to much
larger data sets than those seen elsewhere in the literature. The more
flexible framework is achieved through introducing novel
parameterizations of the model, giving varying degrees of parsimony,
using exponential family models whose structure may be exploited in
various theoretical and algorithmic ways. The algorithms are based on
variational generalized EM algorithms, where the E-steps are augmented
by a minorization-maximization (MM) idea. The bootstrapped standard
error estimates are based on an efficient Monte Carlo network
simulation idea. Last, we demonstrate the usefulness of the
model-based clustering framework by applying it to a discrete-valued
network with more than 131,000 nodes and 17 billion edge variables.
\end{abstract}

%

\end{frontmatter}

\section{Introduction}
\label{secintroduction}

According to Fisher [(\citeyear{Fi22}), page 311],
``the object of statistical methods is the reduction of data.''
The reduction of data is imperative in the case of discrete-valued
networks that may
have hundreds of thousands of nodes and billions of edge variables.
The collection of such large networks is becoming more and more common,
thanks to electronic devices such as cameras and computers.
Of special interest is the identification of influential subsets of
nodes and high-density regions of the network with an eye to break down
the large network into smaller, more manageable components.
These smaller, more manageable components may be studied by more
advanced statistical models, such as advanced exponential family models
[e.g., \citet{FoSd86,StIk90,WsPp96,SnPaRoHa04,HuHa04}].

An example is given by signed networks,
such as trust networks,
which arise in World Wide Web applications.
Users of internet-based exchange networks are invited to classify other
users as either $-1$ (untrustworthy) or $+1$ (trustworthy).
Trust networks can be used to protect users and enhance collaboration
among users [\citet{slashdot2009,PaPa07}].
A second example is the spread of infectious disease through
populations by way of contacts among individuals [\citet{BrNe02,GrWeHu10}].
In such applications,
it may be of interest to identify potential super-spreaders---that is,
individuals who are in contact with many other individuals and who
could therefore
spread the disease to many others---and dense regions of the network
through which disease could spread rapidly.

The current article advances the model-based clustering of large
networks in at least four ways. First, we introduce a simple and
flexible statistical framework for parameterizing models based on
statistical exponential families [e.g., \citet{BN78}] that advances
existing model-based clustering techniques. Model-based clustering of
networks was pioneered by \citet{StNk97}. The simple, unconstrained
parameterizations employed by \citet{StNk97} and others [e.g.,
\citet{NkSt01,ABFX08,Daudin08,ZaPiMiAm10,MaRoCo10}] make sense when
networks are small, undirected and binary, and when there are no
covariates. In general, though, such parameterizations may be
unappealing from both a scientific point of view and a statistical
point of view, as they may result in nonparsimonious models with
hundreds or thousands of parameters. An important advantage of the
statistical framework we introduce here is that it gives researchers a
choice: they can choose interesting features of the data, specify a
model capturing those features, and cluster nodes based on the
specified model. The resulting models are therefore both parsimonious
and scientifically interesting.

Second,
we introduce approximate maximum likelihood estimates of parameters
based on novel variational generalized EM (GEM) algorithms,
which take advantage of minorization-maximization (MM) algorithms
[\citet{HuLa04}] and have computational advantages.
For unconstrained models, tests suggest that the variational GEM
algorithms we propose
can converge quicker and better avoid local maxima than alternative algorithms;
see Sections \ref{seccomparison} and \ref{secapp}.\vadjust{\goodbreak}
In the presence of parameter constraints,
we facilitate computations by exploiting the properties of exponential
families [e.g., \citet{BN78}].
In addition,
we sketch how the variational GEM algorithm can be extended to obtain
approximate Bayesian estimates.

Third,
we introduce bootstrap standard errors to quantify the uncertainty
about the approximate maximum likelihood estimates of the parameters,
whereas other work
has ignored the uncertainty about the approximate maximum likelihood estimates.
To facilitate these bootstrap procedures,
we introduce Monte Carlo simulation algorithms that generate sparse
networks in much less time
than conventional Monte Carlo simulation algorithms.
In fact, without the more efficient Monte Carlo simulation algorithms,
obtaining bootstrap standard errors would be infeasible.

Finally, while model-based clustering has been limited to networks with
fewer than 13,000 nodes and 85 million edge variables
[see the largest data set handled to date, \citet{ZaPiMiAm10}],
we demonstrate that we can handle much larger, nonbinary networks
by considering an internet-based data set
with more than 131,000 nodes and 17 billion edge variables, where
``edge variables''
comprise all observations, including node pairs between which no edge exists.
Many internet-based companies and websites,
such as \url{http://amazon.com}, \url{http://netflix.com} and \url
{http://epinions.com},
allow users to review products and services.
Because most users of the World Wide Web do not know each other and
thus cannot be sure whether to trust each other,
readers of reviews may be interested in an indication of the
trustworthiness of
the reviewers themselves.
A convenient and inexpensive approach is based on evaluations of
reviewers by readers.
The data set we analyze in Section \ref{secapp} comes from
the website \url{http://epinions.com}, which collects such data by
allowing any user $i$ to evaluate any other user $j$ as either untrustworthy,
coded as $y_{ij} = -1$,
or trustworthy,
coded as $y_{ij} = +1$,
where $y_{ij} = 0$ means that user $i$ did not evaluate user $j$
[\citet{PaPa07}].
The resulting network consists of $n = 131$,827 users and $N=n(n-1)=
17$,378,226,102
observations. Since each user can only review a relatively small number
of other users,
the network is sparse: the vast majority of the observations $y_{ij}$
are zero, with only
840,798 negative and positive evaluations.
Our modeling goal, broadly speaking, is both to cluster the users based
on the patterns of
trusts and distrusts in this network and to understand the features of the
various clusters by examining model parameters.

The rest of the article is structured as follows:
A scalable model-based clustering framework based on finite mixture
models is introduced in Section \ref{secmodel}.
Approximate maximum likelihood and Bayesian estimation are discussed in
Sections \ref{secmle} and \ref{secbay}, respectively,
and an algorithm for Monte Carlo simulation of large networks is
described in Section \ref{secsim}.
Section \ref{seccomparison} compares the variational GEM algorithm to
the variational EM algorithm of \citet{Daudin08}.
Section \ref{secapp} applies our methods to the trust network
discussed above.

\section{Models for large, discrete-valued networks}
\label{secmodel}

We consider $n$ nodes,
indexed by integers $1,\ldots, n$,
and edges $y_{ij}$ between pairs of nodes $i$ and $j$,
where $y_{ij}$ can take values in a finite set of $M$ elements.
By convention, $y_{ii} = 0$ for all~$i$,
where $0$ signifies ``no relationship.''
We call the set of all edges $y_{ij}$ a discrete-valued network,
which we denote
by $\mathbf{y}$, and we let $\mY$ denote the set of possible values of~$\mathbf{y}$.
Special cases of interest are
(a) undirected binary networks $\mathbf{y}$,
where $y_{ij} \in\{0, 1\}$ is subject to the linear constraint $y_{ij}
= y_{ji}$ for all $i < j$;
(b) directed binary networks $\mathbf{y}$,
where $y_{ij} \in\{0, 1\}$ for all $i, j$;
and
(c) directed signed networks $\mathbf{y}$,
where $y_{ij} \in\{-1, 0, 1\}$ for all $i, j$.

A general approach to modeling discrete-valued networks is based on
exponential families of distributions [\citet{Bj74,FoSd86}]:
%
\begin{equation}
\label{ergm} P_{\btheta}(\bY= \mathbf{y}\given\bx) = \exp\bigl[
\btheta^\top\bg(\bx, \mathbf{y}) - \psi(\btheta)\bigr],\qquad \mathbf{y}\in\mY,
\end{equation}
where $\btheta$ is the vector of canonical parameters
and $\bg(\bx, \mathbf{y})$ is the vector of canonical statistics
depending on a matrix $\bx$ of covariates,
measured on the nodes or the pairs of nodes,
and the network $\mathbf{y}$,
and $\psi(\btheta)$ is given by
%
\begin{equation}
\psi(\btheta) = \log\sum_{\mathbf{y}^\prime\in\mY} \exp\bigl[\btheta
^\top\bg\bigl(\bx, \mathbf{y}^\prime\bigr)\bigr],\qquad \btheta\in
\mathbb{R}^p,
\end{equation}
and ensures that $P_{\btheta}(\bY= \mathbf{y}\given\bx)$ sums to $1$.

A number of exponential family models have been proposed [e.g.,
\citet{HpLs81,FoSd86,WsPp96,SnPaRoHa04,HuHa04}]. In general, though,
exponential family models are not scalable: the computing time to
evaluate the likelihood function is $\exp(N \log M)$, where $N = n(n -
1)/2$ in the case of undirected edges and $N = n(n - 1)$ in the case of
directed edges, which necessitates time-consuming estimation algorithms
[e.g., \citet{Sn02,HuHa04,moller-etal-2006a,KoRoPa09,CaFr09}].

We therefore restrict attention to scalable exponential family models,
which are characterized by dyadic independence:
%
\begin{equation}
\label{dyadindependence} P_{\btheta}(\bY= \mathbf{y}\given\bx) = \prod
_{i < j}^n P_{\btheta}(D_{ij} =
d_{ij} \given\bx),
\end{equation}
where $D_{ij} \equiv D_{ij}(\bY)$ corresponds to $Y_{ij}$ in the case
of undirected edges and $(Y_{ij}, Y_{ji})$ in the case of directed edges.
The subscripted $i<j$ and superscripted $n$ mean that the product
in (\ref{dyadindependence}) should be taken over
all pairs $(i,j)$ with $1\le i<j\le n$; the same is true for sums as in
(\ref{LB}).

Dyadic independence has at least three advantages:
(a) it facilitates estimation,
because the computing time to evaluate the likelihood function scales
linearly with~$N$;
(b) it facilitates simulation,
because dyads are independent;
and (c) by design it bypasses the so-called model degeneracy problem:
if $N$ is large, some exponential family models without dyadic
independence tend to be ill-defined and impractical
for modeling networks [\citet{St86,Ha03,Sc09b}].

A disadvantage is that most exponential families with dyadic
independence are either simplistic [e.g., models with
identically distributed edges, \citet{ErRe59,gilbert1959}] or
nonparsimonious [e.g., the $p_1$ model with $O(n)$
parameters, \citet{HpLs81}].

We therefore assume that the probability mass function has a
$K$-component mixture form as follows:
%
\begin{eqnarray}
\label{mixturemodel} P_{\bgamma,\btheta}(\bY= \mathbf{y}\given\bx) & = &
\sum
_{\mathbf{z}\in\mZ} P_{\btheta}(\bY= \mathbf{y}\given\bx, \bZ= \mathbf{z})
P_{\bgamma}(\bZ= \mathbf{z})
\nonumber\\[-8pt]\\[-8pt]
& = & \sum_{\mathbf{z}\in\mZ} \prod_{i < j}^n
P_{\btheta
}(D_{ij} = d_{ij} \given\bx, \bZ= \mathbf{z})
P_{\bgamma}(\bZ= \mathbf{z}),
\nonumber
\end{eqnarray}
where $\bZ$ denotes the membership indicators $\bZ_1,\ldots, \bZ_n$
with distributions
%
\begin{equation}
\bZ_i \given\gamma_1,\ldots, \gamma_K
\msim{\mathrm{i.i.d.}} \operatorname{Multinomial}(1; \gamma_1,\ldots,
\gamma_K)
\end{equation}
and $\mZ$ denotes the support of $\bZ$. In some applications, it may be
desired to model the membership indicators $\bZ_i$ as functions of
$\bx$ by using multinomial logit or probit models with $\bZ_i$ as the
outcome variables and $\bx$ as predictors [e.g., \citet{Ta05}]. We
do not elaborate on such models here, but the variational GEM
algorithms discussed in Sections \ref{secmle} and \ref{secbay} could be
adapted to such models.

Mixture models represent a reasonable compromise between model parsimony
and complexity.
In particular, the assumption of conditional dyadic independence
does \textit{not} imply marginal dyadic independence, which means that the
mixture model of (\ref{mixturemodel})
captures some degree of dependence among the dyads.
We give two specific examples of mixture models below.

\subsection*{Example 1}

The $p_1$ model of \citet{HpLs81} for directed, binary-valued networks
may be modified using a mixture model. The original $p_1$
models the sequence of in-degrees (number of incoming edges of nodes)
and out-degrees
(number of outgoing edges of nodes) as well as reciprociated edges, postulating
that the dyads are independent and that the dyadic probabilities are of
the form
%
\begin{equation}
\label{dyadicprob}\qquad P_{\btheta}(D_{ij} = d_{ij}) = \exp
\bigl[(\alpha_i + \beta_j) y_{ij} + (
\alpha_j + \beta_i) y_{ji} + \rho
y_{ij} y_{ji} - \psi_{ij}(\btheta) \bigr],
\end{equation}
where $\btheta= (\alpha_1,\ldots, \alpha_n, \beta_1,\ldots, \beta
_n, \rho)$ and
$\exp\{-\psi_{ij}(\btheta)\}$ is a normalizing constant.
Following \citet{HpLs81},
the parameters $\alpha_i$ may be interpreted as activity or
productivity parameters,
representing the tendencies of nodes $i$ to ``send'' edges to other nodes;
the parameters $\beta_j$ may be interpreted as attractiveness or
popularity parameters,
representing the tendencies of nodes $j$ to ``receive'' edges from other nodes;
and the parameter $\rho$ may be interpreted as a mutuality or
reciprocity parameter,
representing the tendency of nodes $i$ and $j$ to reciprocate edges.

A drawback of this model is that it requires $2n+1$ parameters. Here,
we show how to
extend it to a mixture model that is applicable to both directed and
undirected networks
as well as discrete-valued networks, that
is much more parsimonious, and that allows identification of
influential nodes.

Observe that the dyadic probabilities of (\ref{dyadicprob})
are of the form
%
\begin{equation}
\label{p1extension1} P_{\btheta}(D_{ij} = d_{ij})
\propto\exp\bigl[\btheta_1^\top\bg_1(d_{ij})
+ \btheta_{2i}^\top\bg_{2i}(d_{ij}) +
\btheta_{2j}^\top\bg_{2j}(d_{ij})\bigr],
\end{equation}
where $\btheta_1 = \rho$ is the reciprocity parameter and $\btheta
_{2i} = (\alpha_i, \beta_i)^\top$ and $\btheta_{2j} = (\alpha_j, \beta
_j)^\top$ are the sending and receiving propensities of nodes $i$ and $j$,
respectively.
The corresponding statistics are the reciprocity indicator
$\bg_1(d_{ij}) = y_{ij} y_{ji}$ and the sending and receiving indicators
$\bg_{2i}(d_{ij})=(y_{ij}, y_{ji})^\top$
and
$\bg_{2j}(d_{ij})=(y_{ji}, y_{ij})^\top$ of nodes $i$ and $j$,
respectively.
A mixture model modification of the $p_1$ model
postulates that, conditional on $\bZ$, the dyadic probabilities are
independent and of the form
%
\begin{eqnarray}
\label{p1extension2}
&&P_{\btheta}(D_{ij} = d_{ij} \given
Z_{ik} = Z_{jl} = 1)\nonumber\\[-8pt]\\[-8pt]
&&\qquad \propto\exp\bigl[\btheta_1^\top
\bg_1(d_{ij}) + \btheta_{2k}^\top
\bg_{2k}(d_{ij}) + \btheta_{2l}^\top
\bg_{2l}(d_{ij})\bigr],\nonumber
\end{eqnarray}
where the parameter vectors $\btheta_{2k}$ and $\btheta_{2l}$ depend
on the components $k$ and $l$ to which the nodes $i$ and $j$ belong,
respectively.
The mixture model version of the $p_1$ model is therefore much more
parsimonious provided $K \ll n$ and was proposed by \citet{SPPQ11} in
the case of undirected, binary-valued networks.
Here,
the probabilities of (\ref{p1extension1}) and (\ref
{p1extension2}) are applicable to both undirected and directed networks
as well as discrete-valued networks,
because the functions $\bg_{1k}$ and $\bg_{2l}$ may be customized to
fit the situation
and may even depend on covariates $\bx$, though we have suppressed this
possibility in the notation.
Finally,
the mixture model version of the $p_1$ model admits model-based
clustering of nodes based on indegrees or outdegrees or both.
A small number of nodes with high indegree or outdegree or both is
considered to be influential:
if the corresponding nodes were to be removed,
the network structure would be impacted.

\subsection*{Example 2}
The mixture model of \citet{NkSt01} assumes that, conditional on
$\bZ
$, the dyads are independent and the conditional dyadic probabilities
are of the form
%
\begin{equation}
\label{nowsni} P_{\bpi}(D_{ij} = d \given Z_{ik} =
Z_{jl} = 1) = \pi_{d;kl}.
\end{equation}
In other words,
conditional on $\bZ$,
the dyad probabilities are constant across dyads and do not depend on
covariates.\vadjust{\goodbreak}
It is straightforward to add covariates by writing the conditional dyad
probabilities in canonical form:
%
\begin{equation}
\label{nowsni2}\qquad P_{\btheta}(D_{ij} = d_{ij} \given\bx,
Z_{ik} = Z_{jl} = 1) \propto\exp\bigl[\btheta_1^\top
\bg_1(\bx, d_{ij}) + \btheta_{kl}^\top
\bg_2(\bx, d_{ij}) \bigr],
\end{equation}
where the canonical statistic vectors $\bg_1(\bx, d_{ij})$ and $\bg
_2(\bx, d_{ij})$
may depend on the covariates $\bx$.
If the canonical parameter vectors $\btheta_{kl}$ are constrained by
the linear constraints $\btheta_{kl} = \btheta_k + \btheta_l$, where
$\btheta_k$ and $\btheta_l$ are parameter vectors of the same
dimension as $\btheta_{kl}$,
then the mixture model version of the $p_1$ model arises.
In other words,
the mixture model version of the $p_1$ model can be viewed as a
constrained version of the \citet{NkSt01} model.
While the constrained version can be used to cluster nodes based on degree,
the unconstrained version can be used to identify, for instance,
high-density regions of the network,
corresponding to subsets of nodes with
large numbers of within-subset edges.
These regions may then be studied individually in more detail by using
more advanced statistical models
such as exponential family models without dyadic independence as proposed
by, for example,
\citet{HpLs81}, \citet{FoSd86}, \citet{StIk90},
\citet{WsPp96},
\citet{SnPaRoHa04}
or \citet{HuHa04}.

\subsection*{Other examples}

Other mixture models for networks have been proposed by \citet{Ta05},
\citet{HaRaTa07} and \citet{ABFX08}.
However, these models scale less well to large networks, so we confine
attention here to examples 1 and 2.

\section{Approximate maximum likelihood estimation}
\label{secmle}

A standard approach to maximum likelihood estimation of finite
mixture models is based on the classical EM algorithm,
taking the complete data to be $(\bY, \bZ)$,
where $\bZ$ is unobserved [\citet{DeLaRu77}].
However,
the E-step of an EM algorithm requires the computation of the
conditional expectation of the complete data log-likelihood
function under the distribution of $\bZ\given\bY$,
which is intractable here even in the simplest cases [\citet{Daudin08}].

As an alternative, we consider so-called variational EM algorithms,
which can be considered as generalizations of EM algorithms.
The basic idea of variational EM algorithms is to construct a tractable
lower bound on the intractable log-likelihood function and maximize the
lower bound,
yielding approximate maximum likelihood estimates.
\citet{CeDaLa11} have shown that approximate maximum likelihood
estimators along these lines are---at least in the absence of parameter
constraints---consistent estimators.

We assume that all modeling of $\bY$ can be conditional on covariates
$\bx$ and define
\[
\pi_{d;ij,kl,\bx}(\btheta) = P_{\btheta}(D_{ij} = d \given
Z_{ik} = Z_{jl} = 1, \bx).
\]\eject\noindent
However, for ease of presentation, we drop the notational dependence of
$\pi_{d;ij,kl,\bx}$ on $i, j, \bx$ and make the homogeneity assumption
%
\begin{equation}
\label{homogeneity} \pi_{d;ij,kl,\bx}(\btheta) = \pi_{d;kl}(\btheta)
\qquad\mbox{for all } i, j, \bx,
\end{equation}
which is satisfied by the models in examples 1 and 2.
Exponential parameterizations of $\pi_{d;kl}(\btheta)$,
as in (\ref{dyadicprob}) and (\ref{nowsni2}),
may or may not be convenient.
An attractive property of the variational EM algorithm proposed here is
that it can handle all possible parameterizations of $\pi
_{d;kl}(\btheta)$.
In some cases (e.g., example 1),
exponential parameterizations are more advantageous than others,
while in other cases (e.g., example 2),
the reverse holds.

\subsection{Variational EM algorithm}

Let $A(\mathbf{z}) \equiv P(\bZ=\mathbf{z})$ be an auxiliary distribution
with support $\mZ$.
Using Jensen's inequality,
the log-likelihood function can be bounded below as follows:
%
\begin{eqnarray}
\label{LB{ML}} \log P_{\bgamma,\btheta}(\bY= \mathbf{y}) & = & \log\sum
_{\mathbf{z}\in\mZ} \frac{P_{\bgamma,
\btheta}(\bY= \mathbf{y}, \bZ= \mathbf{z})}{A(\mathbf{z})} A(\mathbf{z})
\nonumber
\\
& \geq& \sum_{\mathbf{z}\in\mZ} \biggl[\log\frac
{P_{\bgamma, \btheta}(\bY= \mathbf{y}, \bZ= \mathbf{z})}{A(\mathbf
{z})}
\biggr] A(\mathbf{z})
\\
& = & E_A\bigl[\log P_{\bgamma, \btheta}(\bY= \mathbf{y}, \bZ= \mathbf{z})
\bigr] - E_A\bigl[\log A(\bZ)\bigr].
\nonumber
\end{eqnarray}
Some choices of $A(\mathbf{z})$ give rise to better lower bounds than others.
To see which choice gives rise to the best lower bound,
observe that the difference between the log-likelihood function and the
lower bound is equal to the Kullback--Leibler divergence from $A(\mathbf
{z})$ to $P_{\bgamma,\btheta}(\bZ= \mathbf{z}\given\bY= \mathbf{y})$:
%
\begin{eqnarray}
\label{kb} & & \log P_{\bgamma,\btheta}(\bY= \mathbf{y}) - \sum
_{\mathbf{z}\in
\mZ} \biggl[\log\frac{P_{\bgamma, \btheta}(\bY= \mathbf{y}, \bZ
= \mathbf{z})}{A(\mathbf{z})} \biggr] A(\mathbf{z})
\nonumber
\\
&&\qquad = \sum_{\mathbf{z}\in\mZ} \bigl[\log P_{\bgamma,\btheta
}(\bY=
\mathbf{y}) \bigr] A(\mathbf{z}) - \sum_{\mathbf{z}\in\mZ} \biggl[\log
\frac{P_{\bgamma, \btheta}(\bY= \mathbf{y}, \bZ= \mathbf
{z})}{A(\mathbf{z})} \biggr] A(\mathbf{z})
\\
&&\qquad = \sum_{\mathbf{z}\in\mZ} \biggl[\log\frac{A(\mathbf
{z})}{P_{\bgamma, \btheta}(\bZ= \mathbf{z}\mid\bY= \mathbf{y})}
\biggr] A(\mathbf{z}).
\nonumber
\end{eqnarray}
If the choice of $A(\mathbf{z})$ were unconstrained in the sense that we
could choose from the set of all distributions with support $\mZ$,
then the best lower bound is obtained by the choice $A(\mathbf{z}) =
P_{\bgamma,\btheta}(\bZ= \mathbf{z}\given\bY= \mathbf{y})$,
which reduces the Kullback--Leibler divergence to $0$ and makes the
lower bound tight.
If the optimal choice is intractable,
as is the case here,
then it is convenient to constrain the choice to a subset of tractable
choices and substitute a choice which, within the subset of tractable
choices, is as close as possible to the optimal choice in terms of
Kullback--Leibler divergence.
A~natural subset of tractable choices is given by introducing the
auxiliary parameters $\balpha= (\balpha_1,\ldots, \balpha_n)$ and setting
%
\begin{equation}
\label{aux} A(\mathbf{z}) = P_{\balpha}(\bZ=\mathbf{z}) = \prod
_{i = 1}^n P_{\balpha_i}(\bZ_i =
\mathbf{z}_i),
\end{equation}
where the marginal auxiliary distributions $P_{\balpha_i}(\bZ_i = \mathbf
{z}_i)$ are Multinomial$(1;\allowbreak \alpha_{i1},\ldots, \alpha_{iK})$.
In this case, the lower bound may be written
%
\begin{eqnarray}
\label{LB} \LB_{\ML}(\bgamma, \btheta; \balpha) &=& E_{\balpha}
\bigl[\log P_{\bgamma, \btheta}(\bY= \mathbf{y}, \bZ= \mathbf{z})\bigr] -
E_{\balpha}
\bigl[\log P_{\balpha}(\bZ)\bigr]
\nonumber\\
&=& \sum_{i < j}^n \sum
_{k=1}^K \sum_{l=1}^K
\alpha_{ik} \alpha_{jl} \log\pi_{d_{ij};kl}(\btheta) \\
&&{}+
\sum_{i=1}^n \sum
_{k=1}^K \alpha_{ik} (\log
\gamma_k - \log\alpha_{ik} ).
\nonumber
\end{eqnarray}
Because equation (\ref{aux}) assumes independence,
the Kullback--Leibler divergence between $P_{\balpha}(\bZ= \mathbf{z})$
and $P_{\bgamma,\btheta}(\bZ= \mathbf{z}\given\bY= \mathbf{y})$,
and thus the tightness of the lower bound, is determined by the
dependence of the random variables $\bZ_1,\ldots, \bZ_n$ conditional
on $\bY$.
If the random variables $\bZ_1,\ldots, \bZ_n$ are independent
conditional on~$\bY$,
then, for each $i$, there exists $\balpha_i$ such that $P_{\balpha
_i}(\bZ_i = \mathbf{z}_i) = P_{\bgamma,\btheta}(\bZ_i = \mathbf{z}_i \mid
\bY= \mathbf{y})$,
which reduces the Kullback--Leibler divergence to $0$ and makes the
lower bound tight.
In general,
the random variables $\bZ_1,\ldots, \bZ_n$ are not independent
conditional on $\bY$ and the Kullback--Leibler divergence (\ref{kb})
is thus positive.

Approximate maximum likelihood estimates of $\bgamma$ and $\btheta$
can be obtained by maximizing the
lower bound in (\ref{LB}) using variational EM algorithms of the
following form, where $t$ is the iteration number:
\begin{longlist}[\textsc{M-step}:]
\item[\textsc{E-step}:]
Letting $\bgamma^{(t)}$ and $\btheta^{(t)}$ denote the current values
of $\bgamma$ and $\btheta$,
maximize $\LB_{\ML}(\bgamma^{(t)}, \btheta^{(t)}; \balpha)$ with
respect to $\balpha$.
Let $\balpha^{(t+1)}$ denote the optimal value of $\balpha$ and
compute $E_{\balpha^{(t+1)}}[\log P_{\bgamma, \btheta}(\bY= \mathbf
{y}, \bZ= \mathbf{z})]$.
\item[\textsc{M-step}:]
Maximize $E_{\balpha^{(t+1)}}[\log P_{\bgamma, \btheta}(\bY= \mathbf
{y}, \bZ= \mathbf{z})]$
with respect to $\bgamma$ and $\btheta$,
which is equivalent to maximizing $\LB_{\ML}(\bgamma, \btheta; \balpha
^{(t+1)})$
with respect to $\bgamma$ and $\btheta$.
\end{longlist}
The method ensures that the lower bound is nondecreasing in the
iteration number:
%
\begin{eqnarray}
\label{nondecreasing1} \LB_{\ML}\bigl(\bgamma^{(t)},
\btheta^{(t)}; \balpha^{(t)}\bigr) & \leq& \LB_{\ML}\bigl(
\bgamma^{(t)}, \btheta^{(t)}; \balpha^{(t+1)}\bigr)
\\
\label{nondecreasing2} & \leq& \LB_{\ML}\bigl(\bgamma^{(t+1)},
\btheta^{(t+1)}; \balpha^{(t+1)}\bigr),
\end{eqnarray}
where inequalities (\ref{nondecreasing1}) and (\ref{nondecreasing2})
follow from the E-step and M-step, respectively.

It is instructive to compare the variational EM algorithm to the
classical EM
algorithm as applied to finite mixture models.
The E-step of the variational EM algorithm minimizes the
Kullback--Leibler divergence between $A(\mathbf{z})$ and
$P_{\bgamma^{(t)},\btheta^{(t)}}(\bZ= \mathbf{z}\given\bY= \mathbf{y})$.
If the choice of $A(\mathbf{z})$ were unconstrained,
then the optimal choice would be $A(\mathbf{z}) = P_{\bgamma
^{(t)},\btheta^{(t)}}(\bZ= \mathbf{z}\given\bY= \mathbf{y})$.
Therefore,
in the unconstrained case,
the E-step of the variational EM algorithm reduces to the E-step of the
classical EM algorithm, so
the classical EM algorithm can be considered to be the optimal
variational EM algorithm.

\subsubsection{Generalized E-step: An MM algorithm}
\label{secmm}

To implement the E-step,
we exploit the fact that the lower bound is nondecreasing as long as
the E-step and M-step increase the lower bound.
In other words,
we do not need to maximize the lower bound in the E-step and M-step.
Indeed, increasing rather than maximizing the lower bound in the E-step
and M-step may have computational advantages when $n$ is large.
In the literature on EM algorithms,
the advantages of incremental E-steps and incremental M-steps are
discussed by \citet{NeHi93} and \citet{DeLaRu77},
respectively.
We refer to the variational EM algorithm with either an incremental
E-step or an incremental M-step or both as a variational generalized
EM, or variational GEM, algorithm.

Direct maximization of $\LB_{\ML}(\bgamma^{(t)}, \btheta^{(t)}; \balpha)$
is unattractive:
equation (\ref{LB}) shows that the lower bound
depends on the products $\alpha_{ik} \alpha_{jl}$ and, therefore,
fixed-point updates of $\alpha_{ik}$ along the lines of
[\citet{Daudin08}] depend on all other $\alpha_{jl}$.
We demonstrate in Section \ref{seccomparison} that the variational EM
algorithm with the fixed-point implementation of the E-step
can be inferior to the variational GEM algorithm when $K$ is large.

To separate\vspace*{1pt} the parameters of the maximization problem,
we increase $\LB_{\ML}(\bgamma^{(t)}, \btheta^{(t)}; \balpha)$ via an
MM algorithm [\citet{HuLa04}].
MM algorithms can be viewed as generalizations of EM algorithms
[\citet{HuLa04}] and are based on iteratively constructing and then
optimizing surrogate (minorizing) functions to facilitate the
maximization problem in certain situations.
We consider here the surrogate function
%
\begin{eqnarray}
\label{qdefn}\quad Q_{\ML}\bigl(\bgamma^{(t)}, \btheta^{(t)};
\balpha^{(t)}, \bolds\balpha\bigr) & = & \sum_{i < j}^n
\sum_{k = 1}^K \sum
_{l = 1}^K \biggl(\alpha_{ik}^2
\frac{\alpha_{jl}^{(t)}}{2 \alpha_{ik}^{(t)}} + \alpha_{jl}^2 \frac
{\alpha_{ik}^{(t)}}{2 \alpha_{jl}^{(t)}} \biggr)
\log\pi_{d_{ij};kl}\bigl(\btheta^{(t)}\bigr)
\nonumber\\[-8pt]\\[-8pt]
&&{} + \sum_{i = 1}^n \sum
_{k = 1}^K \alpha_{ik} \biggl(\log
\gamma_k^{(t)} - \log\alpha_{ik}^{(t)} -
\frac{\alpha_{ik}}{\alpha
_{ik}^{(t)}} + 1 \biggr),
\nonumber
\end{eqnarray}
which we show in Appendix \ref{minorizer} to have the following two properties:
%
\begin{eqnarray}
\label{P1} Q_{\ML}\bigl(\bgamma^{(t)}, \btheta^{(t)},
\balpha^{(t)}; \balpha\bigr) &\le& \LB_{\ML}\bigl(
\bgamma^{(t)}, \btheta^{(t)}; \balpha\bigr) \qquad\mbox{for all }
\balpha,
\\
\label{P2} Q_{\ML}\bigl(\bgamma^{(t)}, \btheta^{(t)},
\balpha^{(t)}; \balpha^{(t)}\bigr) & = & \LB_{\ML}\bigl(
\bgamma^{(t)}, \btheta^{(t)}; \balpha^{(t)}\bigr).
\end{eqnarray}
In the language of MM algorithms, conditions (\ref{P1}) and (\ref
{P2}) establish that
$Q_{\ML}(\bgamma^{(t)}, \btheta^{(t)}, \balpha^{(t)}; \balpha)$ is a
\textit{minorizer} of $\LB_{\ML}(\bgamma^{(t)}, \btheta^{(t)}; \balpha)$
at $\balpha^{(t)}$.
The theory of MM algorithms implies that maximizing the minorizer
with respect to
$\balpha$ forces $\LB_{\ML}(\bgamma^{(t)}, \btheta^{(t)}; \balpha)$
uphill [\citet{HuLa04}].
This maximization, involving $n$ separate quadratic programming
problems of K variables $\balpha_i$ under the constraints
$\alpha_{ik} \geq0$ for all $k$ and $\sum_{k=1}^K \alpha_{ik}=1$,
may be accomplished quickly using the method
described by \citet{Stefanov2004}.
When $n$ is large, it is much easier to update $\balpha$ by
maximizing the $Q_{\ML}$ function, which is the sum of functions of the
individual $\balpha_{i}$, than by maximizing the $\LB_{\ML}$ function,
in which the $\balpha$ parameters
are not separated in this way.
We therefore arrive at the following replacement for the E-step:
\begin{longlist}[\textsc{Generalized E-step}:]
\item[\textsc{Generalized E-step}:] For $i = 1,\ldots, n$, increase
$Q_{\ML}(\bgamma^{(t)}, \btheta^{(t)}, \balpha^{(t)}; \balpha)$
as a function of $\balpha_i$ subject to $\alpha_{ik} \geq0$ for all
$k$ and $\sum_{k = 1}^K \alpha_{ik} = 1$.
Let $\balpha^{(t+1)}$ denote the new value of $\balpha$.
\end{longlist}

\subsubsection{More on the M-step}
\label{secmstep}
To maximize
$\LB_{\ML}(\bgamma, \btheta; \balpha^{(t+1)})$ in the M-step,
examination of (\ref{LB}) shows that maximization
with respect
to $\bgamma$ and $\btheta$ may be accomplished separately.
In fact, for $\bgamma$, there is a simple, closed-form solution:
%
\begin{equation}
\label{gammak} \gamma_k^{(t+1)} = \frac{1}{n} \sum
_{i=1}^n \alpha_{ik}^{(t+1)},\qquad
k = 1,\ldots, K.
\end{equation}
Concerning $\btheta$,
if there are no constraints on $\bpi(\btheta)$ other than $\sum_{d
\in\mD} \pi_{d;kl}(\btheta) = 1$,
it is preferable to maximize with respect to $\bpi= \bpi(\btheta)$
rather than $\btheta$,
because there are closed-form expressions for $\bpi^{(t+1)}$ but not
for $\btheta^{(t+1)}$.
Maximization with respect to $\bpi$ is accomplished by setting
%
\begin{equation}
\label{piupdate}\qquad \pi_{d;kl}^{(t+1)} = \frac{\sum_{i < j}^n \alpha
_{ik}^{(t+1)} \alpha_{jl}^{(t+1)} I(D_{ij} = d)}{\sum_{i <
j}^n \alpha_{ik}^{(t+1)} \alpha_{jl}^{(t+1)}},\qquad d \in
\mD, k, l = 1,\ldots, K.
\end{equation}

If the homogeneity assumption (\ref{homogeneity}) does not hold,
then closed-form expressions for $\bpi$ may not be available.
In some cases, as
in the presence of categorical covariates,
closed form expressions for $\bpi$ are available,
but the dimension of $\bpi$, and thus computing time, increases with
the number of categories.

If equations (\ref{ergm}) and (\ref{dyadindependence}) hold, then
the exponential parametrization $\bpi(\btheta)$ may be inverted to
obtain an approximate maximum likelihood estimate of $\btheta$ after
the approximate MLE of $\bpi$ is found using the variational GEM algorithm.
One method for accomplishing this inversion
exploits the convex duality of exponential families
[\citet{BN78,WaJo08}] and is explained in Appendix \ref{convexduality}.

If, in addition to the constraint $\sum_{d \in\mD} \pi_{d;kl}(\btheta)
= 1$, additional constraints on $\bpi$ are\vadjust{\goodbreak} present, the maximization
with respect to $\bpi$ may either decrease or increase computing time.
Linear constraints on $\bpi$ can be enforced by Lagrange multipliers
and reduce the dimension of $\bpi$ and thus computing time. Nonlinear
constraints on~$\bpi$, as in example 1, may not admit closed form
updates of $\bpi$ and thus may require iterative methods. If so, and if
the nonlinear constraints stem from exponential family
parameterizations of $\bpi(\btheta)$ with natural parameter vector
$\btheta$ as in example~1, then it is convenient to translate the
constrained maximization problem into an unconstrained problem by
maximizing $\LB_{\ML}(\bgamma, \btheta; \balpha^{(t+1)})$ with respect to
$\btheta$ and exploiting the fact that $\LB_{\ML}(\bgamma, \btheta;
\balpha^{(t+1)})$ is a concave function of $\btheta$ owing to the
exponential family membership of $\pi_{d;kl}(\btheta)$
[\citet{BN78}, page~150]. We show in Appendix
\ref{gradienthessian} how the exponential family parameterization can
be used to derive the gradient and Hessian of the lower bound of
$\LB_{\ML}(\bgamma, \btheta; \balpha^{(t+1)})$ with respect to $\btheta$,
which we exploit in Section \ref{secapp} using a Newton--Raphson
algorithm.

\subsection{Standard errors}
\label{secstderr}

Although we maximize the lower bound $\LB_{\ML}(\bgamma,\break \btheta;
\balpha)$ of the log-likelihood function to obtain approximate maximum
likelihood estimates, standard errors of the approximate maximum
likelihood estimates $\hat{\bgamma}$ and $\hat{\btheta}$ based on the
curvature of the lower bound $\LB_{\ML}(\bgamma, \btheta; \balpha)$
may be too small.
The reason is that even when the lower bound is close to the
log-likelihood function,
the lower bound may be more curved than the log-likelihood function
[\citet{wang2005}]; indeed,
the higher curvature helps ensure that $\LB_{\ML}(\bgamma, \btheta;
\balpha)$ is a lower bound of the log-likelihood function $\log
P_{\bgamma,\btheta}(\bY= \mathbf{y})$
in the first place.
As an alternative,
we approximate the standard errors of the approximate maximum
likelihood estimates of $\bgamma$ and $\btheta$ by a parametric
bootstrap method
[\citet{Ef79}] that can be described as follows:
\begin{longlist}[(2)]
\item[(1)] Given the approximate maximum likelihood estimates of
$\bgamma$ and $\btheta$,
sample $B$ data sets.
\item[(2)] For each data set, compute the approximate maximum
likelihood estimates of $\bgamma$ and $\btheta$.
\end{longlist}
In addition to fast maximum likelihood algorithms,
the parametric bootstrap method requires fast simulation algorithms.
We propose such an algorithm in Section \ref{secsim}.

\subsection{Starting and stopping}
\label{secstartingstopping}

As usual with EM-like algorithms, it is a good idea to use multiple
different starting values with the variational EM due to the existence of
distinct local maxima. We find it easiest to use random starts in which
we assign the values of $\balpha^{(0)}$ and then commence with
an M-step. This results in values $\bgamma^{(0)}$ and $\btheta
^{(0)}$, then the algorithm continues with the first E-step, and so on.
The initial $\alpha_{ik}^{(0)}$ are chosen independently uniformly
randomly on $(0,1)$, then each $\balpha_i^{(0)}$ is multiplied by
a normalizing constant chosen so that the elements of $\balpha
_i^{(0)}$ sum to one for every $i$.

The numerical experiments of Section \ref{secapp} use 100 random
restarts each. Ideally, more restarts would be used, yet the size of the
data sets with which we work makes every run somewhat
expensive. We chose the number 100 because we were able to parallelize
on a fairly large scale,
essentially running 100 separate copies of the algorithm. Larger
numbers of runs, such as 1000, would have forced longer run times since
we would have had to run some of the trials
in series rather than in parallel.

As a convergence criterion, we stop the algorithm as soon as
\[
\frac{|\LB_{\ML}(\bgamma^{(t+1)}, \btheta^{(t+1)}; \balpha^{(t+1)}) -
\LB_{\ML}(\bgamma^{(t)}, \btheta^{(t)}; \balpha^{(t)}) | } {
|\LB_{\ML}(\bgamma^{(t+1)}, \btheta^{(t+1)}; \balpha^{(t+1)})|} < 10^{-10}.
\]
We consider the relative change in the objective function
rather than the absolute change or the changes in the parameters
themselves
because (1) even small changes in the parameter values can result in
large changes of the objective function,
and (2) the objective function is a lower bound of the log-likelihood,
so small absolute changes of the objective function may not be worth
the computational effort.

\section{Approximate Bayesian estimation}
\label{secbay}

The key to Bayesian model estimation and model selection is the
marginal likelihood,
defined as
%
\begin{equation}
P(\bY= \mathbf{y}) = \int_{\Gamma} \int_{\Theta}
\sum_{\mathbf
{z}\in\mZ} P_{\bgamma,\btheta}(\bY= \mathbf{y}, \bZ= \mathbf{z})
p(\bgamma, \btheta) \,d\bgamma \,d \btheta,
\end{equation}
where $p(\bgamma, \btheta)$ is the prior distribution of $\bgamma$
and $\btheta$.
To ensure that the marginal likelihood is well-defined,
we assume that the prior distribution is proper,
which is common practice in mixture modeling [\citet{MLPe00},
Chapter 4].
A lower bound on the log marginal likelihood can be derived by
introducing an auxiliary distribution with support $\mZ\times\Gamma
\times\Theta$, where $\Gamma$ is the parameter space of $\bgamma$
and $\Theta$ is the parameter space of $\btheta$.
A natural choice of auxiliary distributions is given by
%
\begin{equation}
A_{\balpha}(\mathbf{z}, \bgamma, \btheta) \equiv\Biggl[\prod
_{i = 1}^n P_{\balpha_{\bZ,i}}(\bZ_i = \mathbf
{z}_i) \Biggr] p_{\balpha_{\bgamma}}(\bgamma) \Biggl[\prod
_{i = 1}^L p_{\balpha_{\btheta}}(\theta_i)
\Biggr],
\end{equation}
where $\balpha$ denotes the set of auxiliary parameters $\balpha_{\bZ
} = (\balpha_{\bZ,1},\ldots, \balpha_{\bZ,n})$, $\balpha_{\bgamma
}$ and~$\balpha_{\btheta}$.

A lower bound on the log marginal likelihood can be derived by Jensen's
inequality:
%
\begin{eqnarray}
\label{LBBayesian}\qquad \log P(\bY= \mathbf{y})& =& \log\int_{\Gamma}
\int_{\Theta} \sum_{\mathbf{z}\in\mZ}
\frac{P_{\bgamma, \btheta}(\bY= \mathbf
{y}, \bZ= \mathbf{z}) p(\bgamma, \btheta)}{A_{\balpha}(\mathbf{z},
\bgamma, \btheta)} A_{\balpha}(\mathbf{z}, \bgamma, \btheta) \,d \bgamma
\,d\btheta
\nonumber\\[-8pt]\\[-8pt]
&\geq& E_{\balpha}\bigl[\log P_{\bgamma, \btheta}(\bY= \mathbf{y}, \bZ= \mathbf
{z}) p(
\bgamma, \btheta)\bigr] - E_{\balpha}\bigl[\log A_{\balpha}(\bZ,
\bgamma, \btheta)\bigr],
\nonumber
\end{eqnarray}
where the expectations are taken with respect to the auxiliary distribution
$A_{\balpha}(\mathbf{z}, \bgamma, \btheta)$.

We denote the right-hand side of (\ref{LBBayesian}) by
$\LB_{B}(\balpha_{\bgamma}, \balpha_{\btheta}; \balpha_{\bZ})$.
By an argument along the lines of (\ref{kb}),
one can show that the difference between the log marginal likelihood
and $\LB_{B}(\balpha_{\bgamma}, \balpha_{\btheta}; \balpha_{\bZ})$
is equal to the Kullback--Leibler divergence from the auxiliary
distribution $A_{\balpha}(\mathbf{z}, \bgamma, \btheta)$ to the
posterior distribution $P(\bZ= \mathbf{z}, \bgamma, \btheta\given\bY=
\mathbf{y})$:
%
\begin{eqnarray}\quad
& & \log P(\bY= \mathbf{y}) - \int_{\Gamma} \int
_{\Theta} \sum_{\mathbf{z}\in\mZ} \biggl[\log
\frac{P_{\bgamma, \btheta}(\bY=
\mathbf{y}, \bZ= \mathbf{z}) p(\bgamma, \btheta)}{A_{\balpha}(\mathbf{z},
\bgamma, \btheta)} \biggr] A_{\balpha}(\mathbf{z}, \bgamma, \btheta)
\,d\bgamma \,d
\btheta
\nonumber\\[-8pt]\\[-8pt]
&&\qquad = \int_{\Gamma} \int_{\Theta} \sum
_{\mathbf{z}\in\mZ} \biggl[\log\frac{A_{\balpha}(\mathbf{z}, \bgamma, \btheta
)}{P(\bZ= \mathbf{z}, \bgamma, \btheta\mid\bY= \mathbf{y})} \biggr] A_{\balpha}(
\mathbf{z}, \bgamma, \btheta) \,d\bgamma \,d\btheta.
\nonumber
\end{eqnarray}

The Kullback--Leibler divergence between the auxiliary distribution and
the posterior distribution can be minimized by a variational GEM
algorithm as follows,
where $t$ is the iteration number:

\begin{longlist}[\textsc{Generalized E-step}:]
\item[\textsc{Generalized E-step}:]
Letting $\balpha_{\bgamma}^{(t)}$ and $\balpha_{\btheta}^{(t)}$
denote the current values of $\balpha_{\bgamma}$ and
$\balpha_{\btheta}$,
increase $\LB_{B}(\balpha_{\bgamma}^{(t)}, \balpha_{\btheta}^{(t)};
\balpha_{\bZ})$
with respect to $\balpha_{\bZ}$.
Let $\balpha_{\bZ}^{(t+1)}$ denote the new value of $\balpha_{\bZ}$.
\item[\textsc{Generalized M-step}:]
Choose new values $\balpha_{\bgamma}^{(t+1)}$
and $\balpha_{\btheta}^{(t+1)}$ that
increase $\LB_{B}(\balpha_{\bgamma}, \balpha_{\btheta}; \balpha
_{\bZ}^{(t+1)})$
with respect to $\balpha_{\bgamma}$ and $\balpha_{\btheta}$.
\end{longlist}
By construction,
iteration $t$ of a variational GEM algorithm increases the lower bound $\LB_{B}
(\balpha_{\bgamma}, \balpha_{\btheta}; \balpha_{\bZ})$:
%
\begin{eqnarray}
\LB_{B}\bigl(\balpha_{\bgamma}^{(t)},
\balpha_{\btheta}^{(t)}; \balpha_{\bZ}^{(t)}\bigr)
& \leq& \LB_{B}\bigl(\balpha_{\bgamma}^{(t)},
\balpha_{\btheta}^{(t)}; \balpha_{\bZ}^{(t+1)}\bigr)
\\
& \leq& \LB_{B}\bigl(\balpha_{\bgamma}^{(t+1)},
\balpha_{\btheta
}^{(t+1)}; \balpha_{\bZ}^{(t+1)}
\bigr).
\end{eqnarray}
A variational GEM algorithm
approximates the marginal likelihood as well as the posterior distribution.
Therefore, it tackles Bayesian model estimation and model selection at
the same time.

Variational GEM algorithms for approximate Bayesian inference are only
slightly more complicated to implement than the variational GEM
algorithms for
approximate maximum likelihood
estimation presented in Section~\ref{secmle}.
To understand the difference, we examine the analogue of (\ref{LB}):
%
\begin{eqnarray}
\label{LBBayesian2}
&&\LB_{B}(\balpha_{\bgamma},
\balpha_{\btheta}; \balpha_{\bZ})\nonumber\\
&&\qquad = \sum
_{i < j}^n \sum_{k = 1}^K
\sum_{l = 1}^K \alpha_{\bZ,ik}
\alpha_{\bZ,jl} E_{\balpha}\bigl[\log\pi_{d_{ij};kl}(\btheta)\bigr]
+ E_{\balpha}\bigl[\log P_{\bgamma}(\bZ= \mathbf{z})\bigr]
\\
&&\qquad\quad{}+ E_{\balpha}\bigl[\log p(\bgamma, \btheta)\bigr] - E_{\balpha}
\bigl[\log A(\bZ= \mathbf{z}, \bgamma, \btheta)\bigr].
\nonumber
\end{eqnarray}
If the prior distributions of $\bgamma$ and $\btheta$ are given by
independent Dirichlet and Gaussian distributions and the auxiliary
distributions of $\bZ_1,\ldots, \bZ_n$, $\bgamma$ and $\btheta$
are given by independent Multinomial, Dirichlet and Gaussian distributions,
respectively,
then the expectations on the right-hand side of (\ref{LBBayesian2})
are tractable,
with the possible exception of the expectations $E_{\balpha}[\log\pi
_{d;kl}(\btheta)]$.
Under the exponential parameterization
%
\begin{equation}
\pi_{d;kl}(\btheta) = \exp\biggl\{\btheta^\top\bg(d) - \log
\sum_{d' \in
\mD} \exp\bigl[ \btheta^\top\bg
\bigl(d'\bigr) \bigr] \biggr\},
\end{equation}
the expectations can be written as
%
\begin{equation}
\label{intractableexp} E_{\balpha}\bigl[\log\pi_{d;kl}(\btheta)
\bigr] = E_{\balpha}[\btheta]^\top\bg(d) - E_{\balpha} \biggl
\{\log\sum_{d' \in\mD} \exp\bigl[ \btheta^\top
\bg\bigl(d'\bigr) \bigr] \biggr\}
\end{equation}
and are intractable.
We are not aware of parameterizations under which the expectations are
tractable.
We therefore use exponential parameterizations and deal with the
intractable nature of the resulting expectations by invoking Jensen's
inequality:
%
\begin{equation}
\label{LBLBBayesian} E_{\balpha}\bigl[\log\pi_{d;kl}(\btheta)\bigr]
\geq E_{\balpha}[\btheta]^\top\bg(d) - \log\sum
_{d'
\in\mD} E_{\balpha} \bigl\{\exp\bigl[\btheta^\top
\bg\bigl(d'\bigr) \bigr] \bigr\}.
\end{equation}
The right-hand side of (\ref{LBLBBayesian}) involves
expectations of independent log-normal random variables, which are tractable.
We thus obtain a looser, yet tractable, lower bound by replacing
$E_{\balpha}[\log\pi_{d;kl}(\btheta)]$ in
(\ref{LBBayesian2}) by the
right-hand side of inequality~(\ref{LBLBBayesian}).

To save space,
we do not address the specific numerical techniques that may be used to
implement the variational GEM algorithm here.
In short,
the generalized E-step is based on an MM algorithm along the lines of
Section \ref{secmm}.
In the generalized M-step,
numerical gradient-based methods may be used.
A~detailed treatment of this Bayesian estimation method and its
implementation, using a more
complicated prior distribution, may be found in \citet{SPPQ11}; code
related to this article is
available at \url{http://sites.stat.psu.edu/\textasciitilde dhunter/code/}.

\section{Monte Carlo simulation}
\label{secsim}

Monte Carlo simulation of large, discrete-valued networks serves at
least three purposes:
\begin{enumerate}[(c)]
\item[(a)] to generate simulated data to be used in simulation studies;
\item[(b)] to approximate standard errors of the approximate maximum
likelihood estimates by parametric bootstrap;
\item[(c)] to assess model goodness of fit by simulation.
\end{enumerate}
A crude Monte Carlo approach is based on sampling $\bZ$ by cycling
through all $n$ nodes
and sampling $D_{ij} \given\bZ$ by cycling through all $n(n-1)/2$ dyads.
However, the running time of such an algorithm is $O(n^2)$,
which is too slow to be useful in practice,
because each of the goals listed above tends to require numerous
simulated data sets.

We propose Monte Carlo simulation algorithms that exploit the fact that
discrete-valued networks tend to be sparse in the sense that one
element of $\mD$
is much more common than all other elements of $\mD$.
An example is given by directed, binary-valued networks,
where $\mD= \{(0, 0), (0, 1), (1, 0), (1, 1)\}$ is the sample space of
dyads and $(0, 0) \in\mD$ tends to dominate all other elements of
$\mD$.

Assume there exists an element $b$ of $\mD$, called the baseline,
that dominates the other elements of $\mD$ in the sense that $\pi
_{b;kl} \gg1 - \pi_{b;kl}$
for all $k$ and $l$.
The Monte Carlo simulation algorithm exploiting the sparsity of large,
discrete-valued networks can be described as follows:
\begin{enumerate}[(2)]
\item[(1)] Sample $\bZ$ by sampling $\mathbf{M} \sim
\operatorname{Multinomial}(n; \gamma_1,\ldots, \gamma_K)$ and assigning nodes $1,\ldots, M_1$ to component $1$, nodes $M_1 + 1,\ldots, M_1 + M_2$ to
component $2$, etc.
\item[(2)] Sample $\bY\given\bZ$ as follows: for each $1\le k\le l
\le K$,
\begin{enumerate}[(c)]
\item[(a)] sample the number of dyads $S_{kl}$ with nonbaseline
values, $S_{kl} \sim\operatorname{Binomial}(N_{kl}, 1 - \pi_{b;kl})$,
where $N_{kl}$ is the number of pairs of nodes belonging to components
$k$ and $l$;
\item[(b)] sample $S_{kl}$ out of $N_{kl}$ pairs of nodes $i<j$
without replacement;
\item[(c)] for each of the $S_{kl}$ sampled pairs of nodes $i < j$,
sample the nonbaseline value $D_{ij}$ according to the probabilities
$\pi_{d;kl}/(1-\pi_{b;kl})$,
$d\in\mD$, $d\ne b$.
\end{enumerate}
\end{enumerate}

In general,
if the degree of any node
(i.e., the number of nonbaseline values for all dyad variables incident
on that node)
has a bounded expectation,
then the expected number of nonbaseline values $S = \sum_{k \leq l} S_{kl}$
in the network
scales with $n$ and the expected running time of the Monte Carlo
simulation algorithm scales with $nK^2|\mD|$.
If $K$ is small and $n$ is large,
then the Monte Carlo approach that exploits the sparsity of large,
discrete-valued networks is superior to the crude Monte Carlo approach.

\section{Comparison of algorithms}
\label{seccomparison}

We compare the variational EM algorithm based on the fixed-point (FP)
implementation of the E-step along the lines of
\citet{Daudin08} to the variational GEM algorithm based on the MM
implementation of the E-step by applying them to
two data sets. The first data set comes from the study on political
blogs by \citet{Adamic2005}.
We convert the binary network of political blogs with two labels,
liberal ($+$1) and conservative ($-$1), into a signed network by assigning
labels of receivers to the corresponding directed edges.
The resulting network has 1490 nodes and 2,218,610 edge variables.
The second data set is the Epinions data set described in Section \ref
{secintroduction} with more than 131,000 nodes and more than 17
billion edge variables.

We compare the two algorithms using the unconstrained network mixture
model of (\ref{nowsni}) with $K = 5$
and $K = 20$ components. For the first data set, we allow up to 1 hour
for $K = 5$ components and up to 6 hours for $K =
20$ components. For the second data set, we allow up to 12 hours for $K
= 5$ components and up to 24 hours for $K = 20$
components. For each data set, for each number of components and for
each algorithm, we carried out 100 runs using
random starting values as described in Section \ref{secstartingstopping}.

\begin{figure}
\begin{tabular}{@{\hspace*{-6pt}}c@{\hspace*{6pt}}c@{}}

\includegraphics{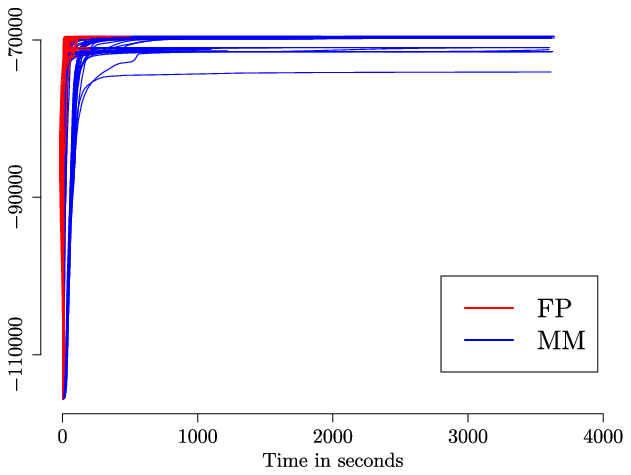}
 & \includegraphics{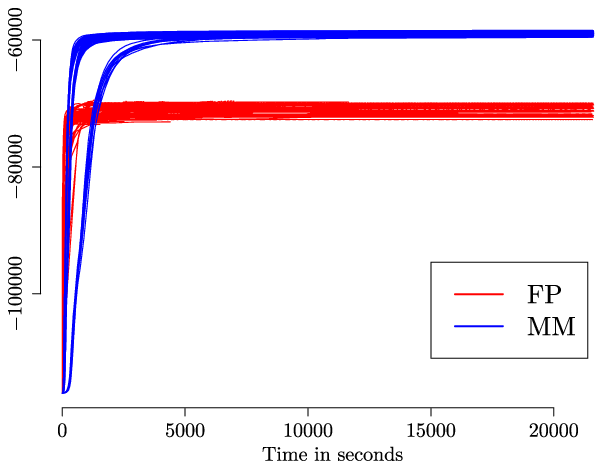}\\
(a) Political blogs data set with $K = 5$ & (b) Political blogs data
set with $K = 20$
\\[6pt]

\includegraphics{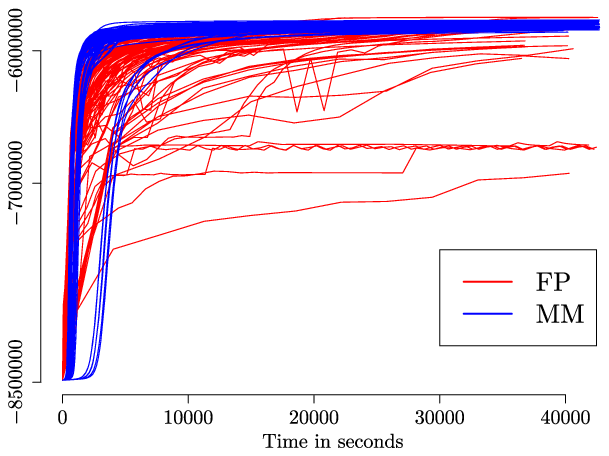}
 & \includegraphics{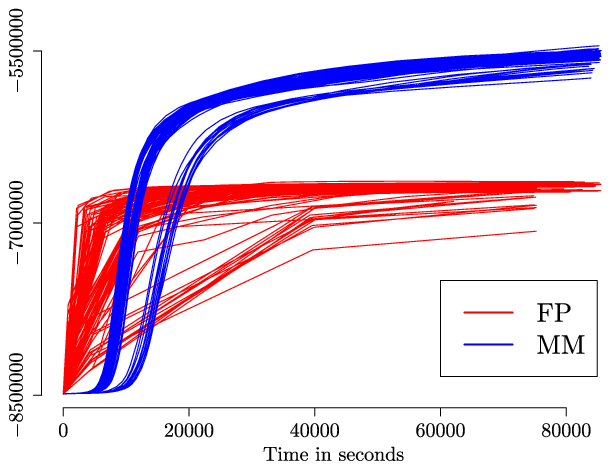}\\
(c) Epinions data set with $K = 5$ & (d) Epinions data set with $K = 20$
\end{tabular}
\caption{Trace plots of the lower bound $\LB_{\ML}(\bgamma^{(t)},
\btheta^{(t)}; \balpha^{(t)})$ of the log-likelihood function for 100
runs each of the variational EM algorithm with FP implementation (red)
and variational GEM algorithm with MM implementation (blue), applied to
the unconstrained network mixture model of (\protect\ref{nowsni}) for
two different data sets.} \label{figFPvsMMConvergenceTraces}
\end{figure}

Figure \ref{figFPvsMMConvergenceTraces} shows
trace plots of the lower bound $\LB_{\ML}(\bgamma^{(t)}, \btheta^{(t)};
\balpha^{(t)})$ of the log-likelihood function,
where red lines refer to the lower bound of the variational EM
algorithm with FP implementation and blue lines refer to
the lower bound of the variational GEM algorithm with MM implementation.
The variational EM algorithm seems to outperform the variational GEM
algorithm in terms of computing time when $K$ and $n$ are small.
However,
when $K$ or $n$ are large,
the variational GEM algorithm appears far superior to the variational
EM algorithm in terms of the lower bounds.
The contrast is most striking when $K$ is large,
though the variational GEM seems to outperform the variational EM
algorithm even when $K$ is small and $n$ is large.
We believe that the superior performance of the variational GEM
algorithm stems from the fact that it separates the parameters
of the maximization problem and reduces the dependence of the updates
of the variational parameters $\alpha_{ik}$, as
discussed in Section \ref{secmm},
while the variational EM algorithm tends to be trapped in local maxima.

Thus, if $K$ and $n$ are small and a computing cluster is available, it
seems preferable to carry out a large number of runs using the
variational EM
algorithm in parallel, using random starting values as described in
Section \ref{secstartingstopping}.
However, if either $K$ or $n$ is large, it is preferable to use the
variational GEM algorithm.
Since the variational GEM algorithm is not prone to be trapped in local maxima,
a small number of long runs may be all that is needed.

\section{Application}
\label{secapp}

Here, we address the problem of clustering the $n=131$,000 users of
the data set
introduced in Section \ref{secintroduction} according to their levels
of trustworthiness,
as indicated by the network of $+$1 and $-1$ ratings given by fellow users.
To this end, we first introduce the individual ``excess trust'' statistics
\[
e_i(\mathbf{y}) = \sum_{1\le j \le n, j\neq i}
y_{ji}.
\]
Since $e_i(\mathbf{y})$ is the number of positive ratings received by user
$i$ in
excess of the number of negative ratings, it is a natural measure
of a user's individual trustworthiness.
Our contention is that consideration of the overall pattern of network
connections results in a more revealing
clustering pattern than a mere consideration of the $e_i(\mathbf{y})$
statistics, and
we support this claim by considering three different clustering methods:
A parsimonious network model using the $e_i(\mathbf{y})$ statistics, the
fully unconstrained
network model of (\ref{nowsni}), and a mixture model that
considers only
the $e_i(\mathbf{y})$ statistics while ignoring the other network structure.

For each method, we assume that the number of categories, $K$, is five.
Partly, this choice is
motivated by the fact that formal model selection methods such as the
ICL criterion
suggested by \citet{Daudin08},
which we discuss in Section 9, suggest dozens if not hundreds of categories,
which complicate summary and interpretation.
Since the reduction of data is the primary task of statistics
[\citet{Fi22}], we want to keep the
number of categories
small and follow the standard practice of internet-based companies and websites,
such as \url{http://amazon.com}
and \url{http://netflix.com}, which use five categories to
classify the trustworthiness of reviewers, sellers and service providers.

Our parsimonious model, which enjoys benefits over the other two
alternatives as we shall see, is based on
%
\begin{eqnarray}
\label{excessDyadModel1}
&&
P_{\btheta}(D_{ij} = d_{ij}
\given Z_{ik} = Z_{jl} = 1)\nonumber \\
&&\qquad \propto \exp\bigl[ \theta^-
\bigl(y_{ij}^- + y_{ji}^-\bigr) + \theta^{+} \bigl(
y_{ij}^+ + y_{ji}^+\bigr) + \theta_k^{\Delta}
y_{ji}
\\
&&\qquad\quad\hspace*{36.5pt}{} + \theta_l^{\Delta} y_{ij} +
\theta^{--} y_{ij}^- y_{ji}^- +
\theta^{++} y_{ij}^+ y_{ji}^+ \bigr],
\nonumber
\end{eqnarray}
where $y_{ij}^-=I(y_{ij}=-1)$ and $y_{ij}^+=I(y_{ij}=1)$ are indicators of
negative and positive edges, respectively.
The parameters in model (\ref{excessDyadModel1}) are not identifiable,
because $y_{ij} = y_{ij}^+ - y_{ij}^-$ and $y_{ji} = y_{ji}^+ - y_{ji}^-$.
We therefore constrain the positive
edge parameter $\theta^+$ to be $0$.
Model (\ref{excessDyadModel1})
assumes in the interest of model parsimony that the propensities to
form negative and positive edges and to reciprocate negative and
positive edges do not vary across clusters; however, the
flexibility afforded by this modeling framework enables us to
define cluster-specific parameters for any of these propensities if we wish.
The conditional probability mass function of the whole network is given by
%
\begin{eqnarray}
\label{excessDyadModel1full}
&& P_{\btheta}(\bY= \mathbf{y}\given\bZ=
\mathbf{z})\nonumber\\
&&\qquad
\propto\exp\Biggl[ \theta^- \sum_{i < j}^n
\bigl(y_{ij}^- + y_{ji}^-\bigr) + \sum
_{k=1}^K \theta_k^{\Delta}
t_k(\mathbf{y}, \mathbf{z})\\
&&\qquad\quad\hspace*{20.2pt}{} + \theta^{--} \sum
_{i < j}^n y_{ij}^- y_{ji}^- +
\theta^{++} \sum_{i < j}^n
y_{ij}^+ y_{ji}^+ \Biggr],\nonumber
\end{eqnarray}
where $t_k(\mathbf{y}, \mathbf{z}) = \sum_{i=1}^n z_{ik}e_i(\mathbf{y})$
is the total excess trust for all nodes in the $k$th category.
The $\theta_k^\Delta$ parameters are therefore measures of the trustworthiness
of each of the categories. Furthermore, these parameters are estimated in
the presence of---that is, after correcting for---the reciprocity
effects as measured
by the parameters
$\theta^{--}$ and $\theta^{++}$, which summarize the overall
tendencies of users to
reciprocate negative and positive ratings, respectively.
Thus, $\theta^{--}$ and $\theta^{++}$ may be considered to measure
overall tendencies toward \textit{lex talionis} and \textit{quid pro quo} behaviors.

One alternative model we consider is the
unconstrained network model obtained from (\ref{nowsni}).
With five components, this model comprises four mixing
parameters $\lambda_1,\ldots, \lambda_4$ in addition to the $\pi
_{d;kl}$ parameters,
of which there are 105: there are nine types of dyads $d$ whenever
$k\ne l$, contributing
$8{5\choose2}=80$\vspace*{1pt} parameters, and six types of dyads $d$ whenever
$k=l$, contributing
an additional $5(5)=25$ parameters.
Despite the increased
flexibility afforded by model (\ref{nowsni}), we view the loss of
interpretability
due to the large number of parameters as a detriment. Furthermore, more
parameters
opens up the possibility of overfitting and, as we discuss below,
appears to make the
lower bound of the log-likelihood function highly multi-modal.

Our other alternative model is a univariate mixture model
applied to the $e_i(\mathbf{y})$ statistics directly, which
assumes that the individual excesses $e_i(\mathbf{y})$
are independent random variables sampled from a distribution with
density
%
\begin{equation}
\label{gaussianmixture} f(x) = \sum_{j=1}^5
\lambda_j \frac1{\sigma_j} \phi\biggl(
\frac{x-\mu_j}{\sigma_j} \biggr),
\end{equation}
where $\lambda_j$, $\mu_j$ and $\sigma_j$ are component-specific
mixing proportions, means and standard deviations, respectively,
and $\phi(\cdot)$ is the standard normal density.
Traditional univariate approaches like this are less
suitable than network-based clustering approaches not only because
by design they neither consider nor inform us about
the topology of the network, which may be relevant,
but also because the individual excesses are not independent:
these $e_i(\mathbf{y})$ are functions of edges, and edges may
be dependent owing to reciprocity (and other forms of dependence not
modeled here),
which decades of research [e.g., \citet{Da68a,HpLs81}] have shown
to be important in shaping social networks.
Unlike the univariate mixture model of (\ref{gaussianmixture}),
the mixture model we employ for networks allows for such dependence.

We use a variational GEM algorithm to estimate the network
model (\ref{excessDyadModel1full}), where the M-step is executed by a
Newton--Raphson algorithm using the gradient and Hessian derived in
Appendix \ref{gradienthessian} with a maximum of 100 iterations. It stops
earlier if the largest absolute value in the gradient vector is less than
$10^{-10}$. By contrast, the unconstrained network model
following from (\ref{nowsni}) employs
a variational GEM algorithm using the exact M-step update (\ref
{piupdate}). The
variational GEM algorithm stops when either the relative change in the
objective function is less than $10^{-10}$ or 6000 iterations are performed.
Most runs require the full 6000 iterations. To
estimate the normal mixture model (\ref{gaussianmixture}), we use the
\texttt{R}
package \texttt{mixtools} [\citet{mixtools}].

\begin{figure}
\begin{tabular}{@{}cc@{}}

\includegraphics{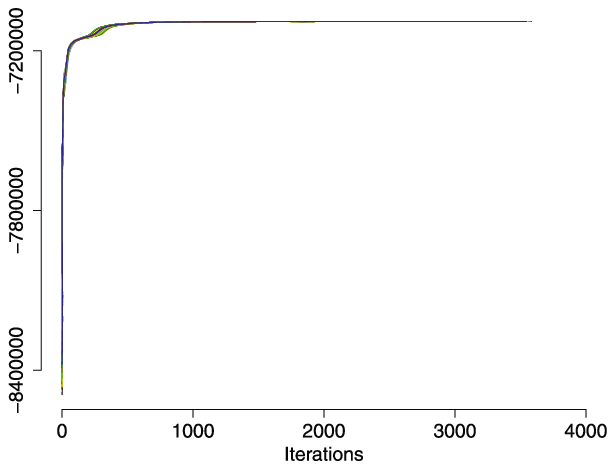}
 & \includegraphics{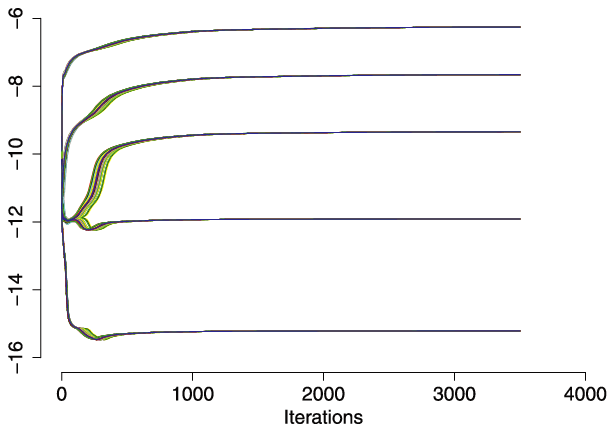}\\
(a) Log-likelihood lower bound & (b) Cluster-specific excess parameters
\end{tabular}
\caption{\textup{(a)} Trace plot of the lower bound $\LB_{\ML}(\bgamma^{(t)},
\btheta^{(t)}; \balpha^{(t)})$ of the log-likelihood function and \textup{(b)}
cluster-specific excess parameters $\theta_k^{\Delta}$, using $100$
runs with random starting values.}
\label{figConvergence}
\end{figure}

To diagnose convergence of the algorithm
for fitting the model (\ref{excessDyadModel1full}),
we present the trace plot of the lower bound of the log-likelihood
function $\LB_{\ML}(\bgamma^{(t)}, \btheta^{(t)};\allowbreak \balpha^{(t)})$ in
Figure \ref{figConvergence}(a) and the trace plot of the cluster-specific
excess parameters $\theta_k^{\Delta}$ in Figure \ref{figConvergence}(b).
Both figures are based on $100$ runs,
where the starting values are obtained by the procedure described in
Section \ref{secstartingstopping}.
The results suggest that all $100$ runs seem to converge to roughly
the same solution.
This fact is somewhat remarkable, since many variational algorithms
appear very sensitive
to their starting values, converging to multiple distinct local optima
[e.g., \citet{DaPiVa10,STMu13}].
For instance, the 100 runs for the unconstrained network model (\ref{nowsni})
produced essentially a unique set of
estimates for each set of random starting values.
Similarly, the normal mixture model algorithm produces many different
local maxima, even after
we try to correct for label-switching by choosing random starting
values fairly tightly clustered
by their mean values.

\begin{figure}
\begin{tabular}{@{}cc@{}}

\includegraphics{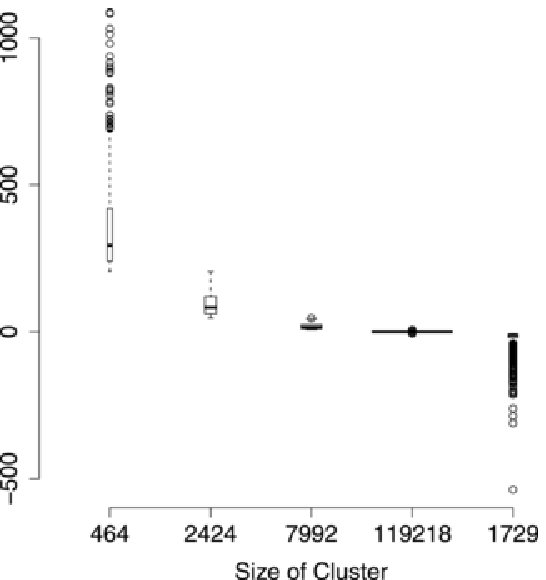}
 & \includegraphics{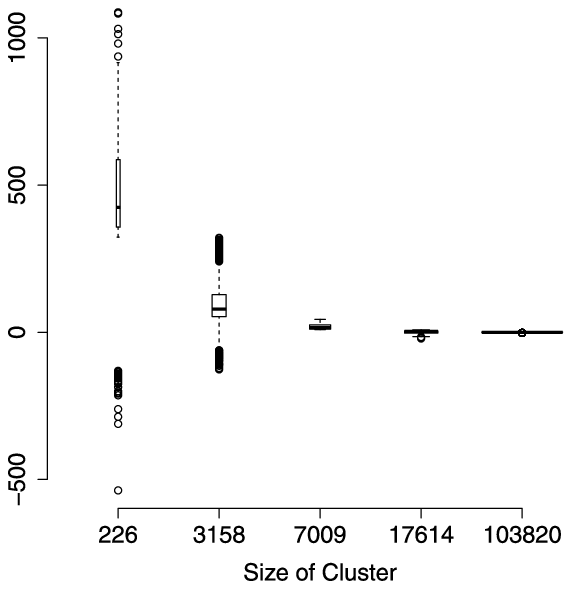}\\
(a) Network mixture model (\ref{excessDyadModel1full}) & (b) Normal
mixture model (\ref{gaussianmixture})\\[6pt]
\multicolumn{2}{@{}c@{}}{
\includegraphics{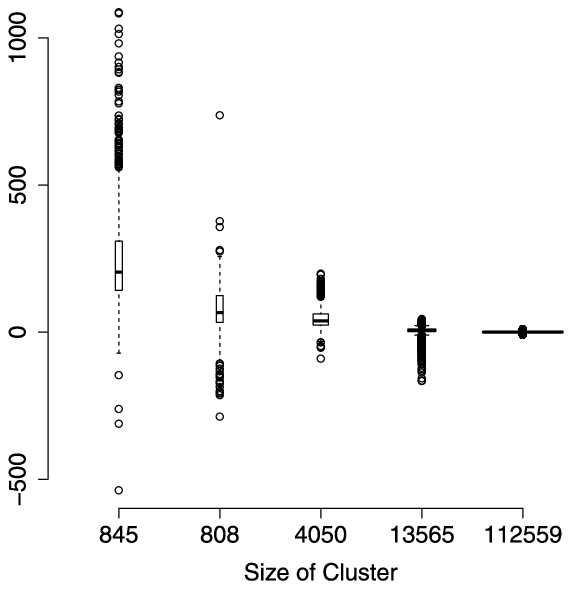}
}\\
\multicolumn{2}{@{}c@{}}{(c) Unconstrained network mixture model (\ref
{nowsni})}
\end{tabular}
\caption{Observed values of excess trust $e_i(\mathbf{y})$,
grouped by highest-probability component of $i$, for \textup{(a)} parsimonious
network mixture model (\protect\ref{excessDyadModel1full}) with 12
parameters, \textup{(b)} normal
mixture model (\protect\ref{gaussianmixture}) with 14 parameters,
and \textup{(c)}
unconstrained
network mixture model (\protect\ref{nowsni}) with 109 parameters.}
\label{figEpinionsExcessTrust}
\end{figure}

Figure \ref{figEpinionsExcessTrust} shows the observed excesses
$e_1(\mathbf{y}),\ldots, e_n(\mathbf{y})$ grouped by clusters
for the best solutions, as measured by likelihood or approximate likelihood,
found for each of the three clustering methods. It appears that
the clustering based on the parsimonious
network model does a better job of separating the $e_i(\mathbf{y})$
statistics into distinct
subgroups---though this is not the sole criterion used---than
the clusterings for the other two models,
which are similar to each other.
In addition, if we use a normal mixture model in which the variances
are restricted to be constant across components, the results are even worse,
with one large cluster and multiple clusters with few nodes.

\begin{figure}
\begin{tabular}{@{}cc@{}}

\includegraphics{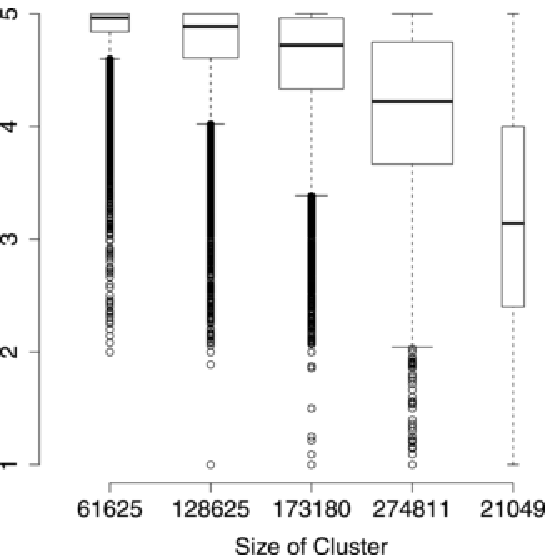}
 & \includegraphics{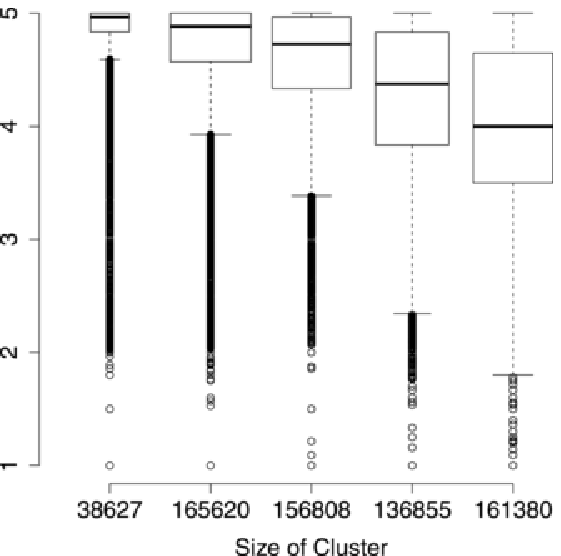}\\
(a) Network mixture model (\ref{excessDyadModel1full}) & (b) Normal
mixture model (\ref{gaussianmixture})\\[6pt]
\multicolumn{2}{@{}c@{}}{
\includegraphics{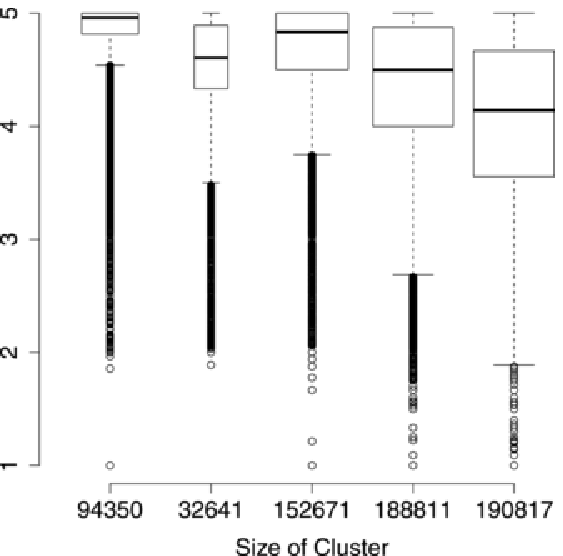}
}\\
\multicolumn{2}{@{}c@{}}{(c) Unconstrained network mixture model
(\ref{nowsni})}
\end{tabular}
\caption{Average ratings of 659,290
articles, grouped according to the highest-probability
category of the article's author,
for \textup{(a)} parsimonious network mixture model (\protect\ref{excessDyadModel1full})
with 12 parameters, \textup{(b)}~normal mixture
model (\protect\ref{gaussianmixture}) with 14 parameters, and \textup{(c)}
unconstrained network mixture model (\protect\ref{nowsni}) with 109
parameters.
The ordering of the five categories, which is the same as in
Figure \protect\ref{figEpinionsExcessTrust},
indicates that the unconstrained network mixture model does not even
preserve the correct ordering of the
median average ratings.}
\label{figEpinionsRatings}
\end{figure}

In Figure \ref{figEpinionsRatings}, we ``ground truth'' the
clustering solutions using external information: the
average ratings of 659,290 articles, grouped according to the
highest-probability category of the article's author.
While in Figure \ref{figEpinionsExcessTrust} the size of each cluster
is the number of users in that
cluster, in Figure \ref{figEpinionsRatings} the size of each cluster is
the number of articles written by users in that cluster.
The widths of the boxes in Figures \ref{figEpinionsExcessTrust}
and \ref{figEpinionsRatings}
are proportional to the square roots of the cluster sizes.

As an objective criterion to compare the three models,
we fit one-way ANOVA models where responses are article ratings and
fixed effects are the group indicators of the articles' authors.
The adjusted R$^2$ values are $0.262$, $0.165$ and $0.172$ for the
network mixture model, the normal
mixture model and the unconstrained network mixture model, respectively.
In other words, the latent structure detected by the 12-component
network mixture
model of (\ref{excessDyadModel1full}) explains the variation
in article ratings better than the 14-parameter univariate mixture
model or the 109-parameter unconstrained network model.

\begin{table}
\caption{95\% Confidence intervals based on parametric bootstrap using
500 simulated networks, with 1000 iterations for each network. The statistic
$\sum_i e_i(\mathbf{y}) Z_{ik}$ equals $\sum_{i} \sum_{j \neq i}
y_{ji} Z_{ik}$, where $Z_{ik} = 1$ if user $i$ is a member of cluster
$k$ and $Z_{ik} = 0$ otherwise}\label{tableepinions}
\begin{tabular*}{\tablewidth}{@{\extracolsep{\fill}}lcd{3.3}c@{}}
\hline
& & \multicolumn{1}{c}{\textbf{Parameter}} & \textbf{Confidence} \\
\textbf{Parameter} & \textbf{Statistic} & \multicolumn{1}{c}{\textbf{estimate}} & \textbf{interval}
\\ \hline
Negative edges ($\theta^-$) & $\sum_{ij} y_{ij}^-$ & -24.020 &
$(-24.029, -24.012)$ \\[3pt]
Positive edges ($\theta^+$) & $\sum_{ij} y_{ij}^+$ & 0 & --- \\[3pt]
Negative reciprocity ($\theta^{--}$)& $\sum_{ij} y_{ij}^- y_{ji}^-$ &
8.660 & $(8.614, 8.699)$ \\[3pt]
Positive reciprocity ($\theta^{++}$)& $\sum_{ij} y_{ij}^+ y_{ji}^+$ &
9.899 & $(9.891, 9.907)$ \\[3pt]
Cluster 1 trustworthiness ($\theta_1^\Delta$) & $\sum_i e_i(\mathbf
{y}) Z_{i1}$ & -6.256 & $(-6.260, -6.251)$ \\[3pt]
Cluster 2 trustworthiness ($\theta_2^\Delta$) & $\sum_i e_i(\mathbf
{y}) Z_{i2}$ & -7.658 & $(-7.662, -7.653)$ \\[3pt]
Cluster 3 trustworthiness ($\theta_3^\Delta$) & $\sum_i e_i(\mathbf
{y}) Z_{i3}$ & -9.343 & $(-9.348, -9.337)$ \\[3pt]
Cluster 4 trustworthiness ($\theta_4^\Delta$) & $\sum_i e_i(\mathbf
{y}) Z_{i4}$ & -11.914 & $(-11.919, -11.908)$ \\[3pt]
Cluster 5 trustworthiness ($\theta_5^\Delta$) & $\sum_i e_i(\mathbf
{y}) Z_{i5}$ & -15.212 & $(-15.225, -15.200)$\\
\hline
\end{tabular*}
\end{table}

\begin{figure}[b]

\includegraphics{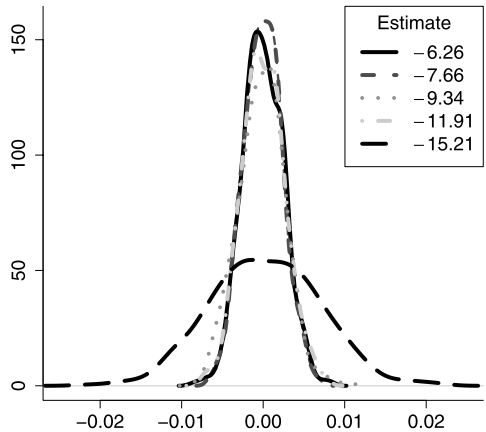}

\caption{Kernel density estimates of the five bootstrap samples
of the trustworthiness parameters, shifted so that each
component's estimated parameter value (shown in the legend) equals
zero.}
\label{figBootstrap1}
\end{figure}

Table \ref{tableepinions} reports estimates of the $\theta$ parameters
from model (\ref{excessDyadModel1full}) along with
95\% confidence intervals reported
in that table obtained by simulating 500 networks using the
method of Section \ref{secsim} and the parameter estimates obtained
via our algorithm. For each network, we run our algorithm for
1000 iterations starting at the M-step, where the $\balpha$
parameters are initialized to reflect the ``true'' component to
which each node is assigned by the simulation algorithm by setting
$\alpha_{ik}=10^{-10}$ for $k$ not equal to the true component and
$\alpha_{ik}=1-4\times10^{-10}$ otherwise. This is done to
eliminate the so-called label-switching problem, which is rooted in
the invariance of the likelihood function to switching the labels of
the $5$ components and which can affect bootstrap samples in the
same way it can affect Markov chain Monte Carlo samples from the
posterior of finite mixture models [\citet{St00}]. The sample 2.5\%
and 97.5\% quantiles form the confidence intervals shown. In
addition, we give density estimates of the five trustworthiness
bootstrap samples in Figure \ref{figBootstrap1}.
%
Table \ref{tableepinions} shows that some clusters of
users are much more trustworthy than others.
In addition,
there is statistically significant evidence that users rate others in
accordance with
both \textit{lex talionis} and \textit{quid pro quo}, since both $\theta
^{--}$ and $\theta^{++}$ are
positive.
These findings suggest that the ratings of pairs of users $i$ and $j$ are,
perhaps unsurprisingly, dependent and not free of self-interest.

Finally, a few remarks concerning the parametric bootstrap are appropriate.
While we are encouraged by the fact that bootstrapping is even
feasible for problems of this size, there are aspects of our
investigation that will need to be addressed with further research.
First, the bootstrapping is so time-consuming that we were forced to
rely on computing clusters with multiple computing nodes to generate
a bootstrap sample in reasonable time. Future work could focus on
more efficient bootstrapping. Some work on efficient bootstrapping
was done by \citet{kleiner2011scalable}, but it is restricted to simple
models and not applicable here.

Second, when the variational GEM algorithm is initialized at random
locations, it may converge to local maxima whose $\LB_{\ML}(\bgamma,
\btheta; \balpha)$ values are inferior to the solutions attained
when the algorithm is initialized at the ``true'' values used to
simulate the networks. While it is not surprising that variational
GEM algorithms converge to local maxima, it is surprising that the
issue shows up in some of the simulated data sets but not in the
observed data set. One possible explanation is that the structure of
the observed data set is clear cut, but that the components of the
estimated model are not sufficiently separated. Therefore, the
estimated model may place nonnegligible probability mass on
networks where two or more subsets of nodes are hard to distinguish
and the variational GEM algorithm may be attracted to local maxima.

Third, some groups of confidence intervals, such as the first four
trustworthiness parameter intervals, have more or less the same width.
We do
not have a fully satisfying explanation for this result; it may be a coincidence
or it may have some deeper cause related to the difficulty of the
computational problem.

In summary,
we find that the clustering framework we introduce here provides
useful results for a very large network. Most importantly, the sensible
application of statistical modeling ideas, which reduces the unconstrained
109-parameter model to a constrained 12-parameter model,
produces vastly superior results in terms of interpretability,
numerical stability and predictive performance.

\section{Discussion}

The model-based clustering framework outlined here represents several advances.
An attention to standard statistical modeling ideas relevant in the
network context
improves model parsimony and interpretability relative to fully unconstrained
clustering models, while also suggesting a viable method for assessing
precision of estimates obtained.
Algorithmically, our advances allow us to apply a variational EM idea, recently
applied to network clustering models in numerous publications
[e.g., \citet{NkSt01,ABFX08,Daudin08,ZaPiMiAm10,MaRoCo10}],
to networks far larger than any that have been considered to date.
We have applied our methods to networks with over a hundred thousand nodes\vadjust{\goodbreak}
and signed edges, indicating how they extend to categorical-valued
edges generally or
models that incorporate other covariate information.
In practice, these methods could have myriad uses, from identifying
high-density regions of large networks to selecting among competing
models for a single network to testing specific network effects of
scientific interest
when clustering is present.

To achieve these advances, we have focused exclusively on models exhibiting
dyadic independence conditional on the cluster memberships of nodes.
It is important to remember that these models are \textit{not}
dyadic independence models overall, since the clustering itself
introduces dependence. However,
to more fully capture network effects such as transitivity, more
complicated
models may be needed, such as the latent space models of
\citet{HpRaHm01}, \citet{ScSn03} or \citet{HaRaTa07}. A
major drawback
of latent space models is that they tend to be less scalable than the models
considered here. An example is given by the variational Bayesian
algorithm developed by \citet{STMu13} to estimate the latent space
model of \citet{HaRaTa07}. The running time of the algorithm is
$O(n^2)$ and it has therefore not been applied to networks with more
than $n = 300$ nodes and $N = 89$,700 edge variables.
An alternative to the variational Bayesian algorithm of \citet{STMu13}
based on case-control
sampling was proposed by \citet{RaNiHoYe12}.
However, while the computing time of this alternative algorithm is $O(n)$,
the suggested preprocessing step, which requires
determining the shortest path length between pairs of nodes, is
$O(n^2)$.
As a result, the largest network \citet{RaNiHoYe12} analyze is an
undirected network
with $n = 2716$ nodes and $N = 3$,686,970 edge variables.

In contrast, the running time of the variational GEM algorithm proposed
here is $O(n)$ in the constrained and $O(f(n))$
in the unconstrained version of the \citet{NkSt01} model, where $f(n)$
is the number of edge variables whose value is
not equal to the baseline value. It is worth noting that $f(n)$ is
$O(n)$ in the case of sparse graphs and, therefore, the
running time of the variational GEM algorithm is $O(n)$ in the case of
sparse graphs. Indeed, even in the presence of
the covariates, the running time of the variational GEM algorithm is
$O(n \prod_{i=1}^I C_i)$ with categorical
covariates, where $I$ is the number of covariates and $C_i$ is the
number of categories of the $i$th covariate. We have
demonstrated that the variational GEM algorithm can be applied to
networks with more than $n = 131$,000 nodes and $N =
17$ billion edge variables.

While the running time of $O(n)$ shows that the variational GEM algorithm
scales well with $n$, in practice, the ``G'' in ``GEM'' is an important
contributor
to the speed of the variational GEM algorithm: merely increasing
the lower bound using an MM algorithm rather than actually maximizing it
using a fixed-point algorithm along the lines of \citet{Daudin08}
appears to
save much computing time for large networks, though an exhaustive
comparison of these two methods is a topic for further
investigation.\looseness=-1

An additional increase in speed might be gained by exploiting
acceleration methods such as quasi-Newton methods [\citet
{PwFbTsVw86}, Section
10.7], which have shown promise in the case of MM
algorithms [\citet{HuLa04}] and which might accelerate the MM
algorithm in the E-step of the variational GEM algorithm. However,
application of these methods is complicated in the current modeling
framework because of the exceptionally large number of auxiliary
parameters introduced
by the variational augmentation.

We have neglected here the problem of selecting the number of clusters.
\citet{Daudin08} propose making this selection
based on the so-called ICL criterion, but it is not known how the ICL
criterion behaves when the intractable incomplete-data log-likelihood
function in the ICL criterion is replaced by a variational-method
lower bound. In our experience,
the magnitude of the changes in
the maximum lower bound value achieved with multiple random
starting parameters is at least as large as the magnitude of the
penalization imposed
on the log-likelihood by the ICL criterion. Thus, we have been
unsuccessful in
obtaining reliable ICL-based results for very large networks.
More investigation
of this question, and of the selection of the number of clusters in general,
seems warranted.

By demonstrating that scientifically interesting clustering models
can be applied to very large networks by extending
the variational-method ideas developed for network data sets
recently in the statistical literature,
we hope to encourage further investigation of the possibilities of these
and related clustering methods.

The source code, written in \texttt{C++}, and data files used in
Sections \ref{seccomparison} and \ref{secapp} are publicly
available at
\url{http://sites.stat.psu.edu/\textasciitilde dhunter/code}.

\begin{appendix}

\section{Obtaining a Minorizer of the lower bound}
\label{minorizer}

The lower bound $\LB_{\ML}(\bgamma, \btheta; \balpha)$ of the
log-likelihood function can be written as
%
\begin{eqnarray}
\label{LBappendix} \LB_{\ML}(\bgamma, \btheta; \balpha) &=& \sum
_{i < j}^n \sum_{k=1}^K
\sum_{l=1}^K \alpha_{ik}
\alpha_{jl} \log\pi_{d_{ij};kl}(\btheta)\nonumber\\[-8pt]\\[-8pt]
&&{} + \sum
_{i=1}^n \sum_{k=1}^K
\alpha_{ik} ( \log\gamma_k - \log\alpha_{ik} ).\nonumber
\end{eqnarray}
Since $\log\pi_{d_{ij};kl}(\btheta) < 0$ for all $\btheta$,
the arithmetic-geometric mean inequality implies that
%
\begin{equation}
\alpha_{ik} \alpha_{jl} \log\pi_{d_{ij};kl}(\btheta) \ge
\biggl(\alpha_{ik}^2 \frac{\hat\alpha_{jl}}{2 \hat\alpha
_{ik}} +
\alpha_{jl}^2 \frac{\hat\alpha_{ik}}{2 \hat\alpha
_{jl}} \biggr) \log
\pi_{d_{ij};kl}(\btheta)\vadjust{\goodbreak}
\end{equation}
[\citet{HuLa04}],
with equality if $\alpha_{ik}=\hat\alpha_{ik}$ and $\alpha
_{jl}=\hat\alpha_{jl}$.
In addition, the concavity of the logarithm function gives
%
\begin{equation}
-\log\alpha_{ik} \ge-\log\hat\alpha_{ik} -
\frac{\alpha
_{ik}}{\hat\alpha_{ik}} + 1
\end{equation}
with equality if $\alpha_{ik}=\hat\alpha_{ik}$.
Therefore,
function $Q_{\ML}(\bgamma, \btheta, \balpha; \hat{\balpha})$ as
defined in (\ref{qdefn}) possesses properties (\ref{P1}) and (\ref{P2}).

\section{Convex duality of exponential families}
\label{convexduality}

We show how closed-form expressions of $\btheta$ in terms of $\bpi$
can be obtained by exploiting the convex duality of exponential
families. Let
%
\begin{equation}
\label{conuugatedual} \psi^*(\bmu) = \sup_{\btheta} \bigl\{
\btheta^\top\bmu- \psi(\btheta) \bigr\}
\end{equation}
be the Legendre--Fenchel transform of $\psi(\btheta)$,
where $\bmu\equiv\bmu(\btheta) = E_{\btheta}[\bg(\bY)]$ is the
mean-value parameter vector and the subscripts $k$ and $l$ have been dropped.
By Barndorff-Nielsen [(\citeyear{BN78}), page 140] and
Wainwright and Jordan [(\citeyear{WaJo08}), pages 67 and 68],
the Legendre--Fenchel transform of $\psi(\btheta)$ is self-inverse
and, thus, $\psi(\btheta)$ can be written as
%
\begin{equation}
\label{dual2} \psi(\btheta) = \sup_{\bmu} \bigl\{ \bolds
\theta^\top\bmu- \psi^*(\bmu) \bigr\} = \sup_{\bpi}
\bigl\{ \bolds\theta^\top\bmu(\bpi) - \psi^*\bigl(\bmu(\bpi)\bigr) \bigr
\},
\end{equation}
where $\bmu(\bpi) = \sum_{d \in\mD} \bg(d) \pi_{d}$ and $\psi
^*(\bmu(\bpi)) = \sum_{d \in\mD} \pi_d \log\pi_d$.
Therefore,
closed-form expressions of $\btheta$ in terms of $\bpi$ may be found
by maximizing $\bolds\theta^\top\bmu(\bpi) - \psi^*(\bmu(\bpi))$
with respect to $\bpi$.

\section{Gradient and Hessian of lower bound}
\label{gradienthessian}

We are interested in the gradient and Hessian with respect
to the parameter vector $\btheta$ of the lower bound in
(\ref{LBappendix}).
The two examples of models considered in Section \ref{secmodel}
assume that the conditional dyad probabilities $\pi
_{d_{ij};kl}(\btheta)$ take the form
%
\begin{equation}
\pi_{d_{ij};kl}(\btheta) = \exp\bigl[\bta_{kl}(
\btheta)^\top\bg(d_{ij}) - \psi_{kl}(\btheta)\bigr],
\end{equation}
where $\bta_{kl}(\btheta) = \bA_{kl} \btheta$ is a linear
function of parameter vector $\btheta$ and $\bA_{kl}$ is a matrix of
suitable order depending on components $k$ and $l$.
It is convenient to absorb the matrix $\bA_{kl}$ into the statistic
vector $\bg(d_{ij})$ and write
%
\begin{equation}
\pi_{d_{ij};kl}(\btheta) = \exp\bigl[\btheta^\top
\bt_{kl}(d_{ij}) - \psi_{kl}(\btheta)\bigr],
\end{equation}
where $\bt_{kl}(d_{ij}) = \bA_{kl}^\top\bg(d_{ij})$.
Thus, we may write
%
\begin{equation}\quad
\LB_{\ML}(\bgamma, \btheta; \balpha) = \sum
_{i < j}^n \sum_{k=1}^K
\sum_{l=1}^K \alpha_{ik}
\alpha_{jl} \bigl[\btheta^\top\bt_{kl}(d_{ij})
- \psi_{kl}(\btheta) \bigr] + \mbox{const},
\end{equation}
where ``const'' denotes terms which do not depend on $\btheta$ and
%
\begin{equation}
\psi_{kl}(\btheta) = \log\sum_{d \in\mD} \exp
\bigl[\btheta^\top\bt_{kl}(d)\bigr].
\end{equation}

Since the lower bound $\LB_{\ML}(\bgamma, \btheta; \balpha)$ is a
weighted sum of exponential family log-probabilities,
it is straightforward to obtain the gradient and Hessian of
$\LB_{\ML}(\bgamma, \btheta; \balpha)$ with respect to $\btheta$,
which are given by
%
\begin{equation}
\nabla_{\theta} \LB_{\ML}(\bgamma, \btheta; \balpha) = \sum
_{i < j}^n \sum_{k=1}^K
\sum_{l=1}^K \alpha_{ik}
\alpha_{jl} \bigl\{\bt_{kl}(d_{ij}) -
E_{\btheta}\bigl[\bt_{kl}(D_{ij})\bigr] \bigr\}
\end{equation}
and
%
\begin{equation}
\nabla_{\theta}^2 \LB_{\ML}(\bgamma, \btheta;
\balpha) = - \sum_{i < j}^n \sum
_{k=1}^K \sum_{l=1}^K
\alpha_{ik} \alpha_{jl} E_{\btheta}\bigl[
\bt_{kl}(D_{ij}) \bt_{kl}(D_{ij})^\top
\bigr],
\end{equation}
respectively.

In other words,
the gradient and Hessian of $\LB_{\ML}(\bgamma, \btheta; \balpha)$
with respect to $\btheta$ are weighted sums of expectations---the
means, variances and covariances of statistics.
Since the sample space of dyads $\mD$ is finite and, more often than
not, small,
these expectations may be computed by complete enumeration of all
possible values of $d \in\mD$
and their probabilities.
\end{appendix}

\section*{Acknowledgments}

We are grateful to Paolo Massa and Kasper Souren of
\href{http://www.trustlet.org}{trustlet.org}
for sharing the \href{http://www.epinions.com}{epinion.com} data


%

\printaddresses

\end{document}